\documentstyle[amsfonts,amssymb,epsfig,12pt]{article}

\makeatletter
\renewcommand \thesection {\@arabic\c@section}
\renewcommand\thesubsection   {\thesection.\@arabic\c@subsection}
\renewcommand\thesubsubsection{\thesubsection .\@arabic\c@subsubsection}
\renewcommand\theparagraph    {\thesubsubsection.\@arabic\c@paragraph}
\renewcommand\section{\@startsection {section}{1}{\z@}%
                                   {-3.5ex \@plus -1ex \@minus -.2ex}%
                                   {1.9ex \@plus.2ex}%
                                   {\normalfont\large\bfseries\centering}}
\renewcommand\subsection{\@startsection{subsection}{2}{\z@}%
                                     {-2ex\@plus -1ex \@minus -.2ex}%
                                     {1.2ex \@plus .2ex}%
                                    {\normalfont\normalsize\bfseries\centering}
}
\renewcommand\subsubsection{\@startsection{subsubsection}{3}{\z@}%
                                     {-2ex\@plus -1ex \@minus -.2ex}%
                                     {.5ex \@plus .2ex}%
                                     {\normalfont\normalsize\em}}
\renewcommand\paragraph{\@startsection{paragraph}{4}{\z@}%
                                    {3.25ex \@plus1ex \@minus.2ex}%
                                    {-1em}%
                                    {\normalfont\normalsize\em}}
\renewcommand\subparagraph{\@startsection{subparagraph}{5}{\parindent}%
                                       {3.25ex \@plus1ex \@minus .2ex}%
                                       {-1em}%
                                      {\normalfont\normalsize\em}}
\makeatother

\newcommand\abs[1]{{\left| #1 \right|}}

\newcommand{\aprl}{\,{\stackrel{<}{{}_\sim}}\,}

\newcounter{subequation}
	\newenvironment{subequation}%
	{\addtocounter{equation}{-1}%
	\stepcounter{subequation}%
	\begin{equation}}%
	{\end{equation}%
}

\newcommand{\beq}{\begin{equation}}
\newcommand{\eeq}{\end{equation}}
\newcommand{\bseq}{\begin{subequation}}
\newcommand{\eseq}{\end{subequation}}
\newcommand{\bea}{\begin{eqnarray}}
\newcommand{\eea}{\end{eqnarray}}

\newcommand{\refeq}[1]{(\ref{#1})}

\newcommand{\supp}{{\mathrm{supp}\ }}

\newcommand{\pmk}{{{z_k}}}

\newcommand{\plumi}{{\scriptstyle{(\pm)}}}

\newcommand{\eps}{\epsilon}
\newcommand{\veps}{\varepsilon} 

\newcommand{\QED}{$\quad$\textrm{Q.E.D.}}

\newcommand{\me}{m_{\mathrm{e}}}

\newcommand{\lC}{\lambda_{\mathrm{C}}}

\newcommand{\bulldif}[1]{{{#1}^{\!\!\!\stackrel{\bullet}{\phantom{.}}}}{}}

\newcommand{\dd}{{\mathrm{d}}}

\newcommand{\cN}{{\cal N}}

\newcommand{\MM}{{\Bbb M}}
\newcommand{\NN}{{\Bbb N}}

\newcommand{\RR}{{\Bbb R}}

\newcommand{\MTWtens}[1]{{\textbf{\textsf{#1}}}}

\newcommand{\MQ}{{\MTWtens{M}}}
\newcommand{\FQ}{{\MTWtens{F}}}

\newcommand{\dQ}{{\MTWtens{d}}}

\newcommand{\MTWvec}[1]{{\mathbf{#1}}}

\newcommand{\AQ}{{\MTWvec{A}}}
\newcommand{\JQ}{{\MTWvec{J}}}

\newcommand{\uQ}{{\MTWvec{u}}}
\newcommand{\UQ}{{\MTWvec{U}}}

\newcommand{\Eta}{{\mathrm{H}}}

\newcommand{\alongHk}{{\big|_{_{\Eta_k}}\big.}}

\newcommand{\SPvec}[1]{{\textbf{\textsl{#1}}}}

\newcommand{\dvol}{{\mathrm{d}^3}}

\newcommand{\Hodge}{{{}^\star}}

\newcommand{\haelfte}{{\textstyle{\frac{1}{2}}}}
\newcommand{\viertel}{{\textstyle{\frac{1}{4}}}}

\newtheorem{defn}{Definition}[section] 

\newtheorem{Coro}[defn]{Corollary}

\newtheorem{Prop}[defn]{Proposition}
\newtheorem{Rema}[defn]{Remark}

\begin{document}

\title{ ELECTROMAGNETIC FIELD THEORY \\
	WITHOUT DIVERGENCE PROBLEMS \\
	2. A Least Invasively Quantized Theory}

\author{\textbf{MICHAEL K.-H. KIESSLING}\\
	Department of Mathematics\\
	Rutgers, The State University of New Jersey\\
	110 Frelinghuysen Rd., Piscataway, NJ 08854}

\date{}

\maketitle
\begin{abstract}
\noindent
 Classical electrodynamics based on the Maxwell-Born-Infeld field equations 
coupled with a Hamilton--Jacobi law of point charge motion is partially quantized. 
 The Hamilton--Jacobi phase function is supplemented by a dynamical amplitude
field on configuration space.
 Both together combine into a single complex 
wave function satisfying a relativistic Klein--Gordon equation that is self-consistently 
coupled to the evolution equations for the point charges and the electromagnetic 
fields.
 Radiation-free stationary states exist.
 The hydrogen spectrum is discussed in some detail.
 Upper bounds for Born's `aether constant' are obtained.
 In the limit of small velocities of and negligible radiation 
from the point charges, the model reduces to Schr\"odinger's equation with 
Coulomb Hamiltonian, coupled with the de Broglie--Bohm guiding equation.
\end{abstract} 

\textit{Electromagnetism}: electromagnetic fields, point charges, wave 
functions;
\textit{Determinism}: Maxwell--Born--Infeld field equations,
de Broglie--Bohm law of  quantum motion,
Klein--Gordon equation;
\textit{Probability}: configuration space, Born's statistical law.
\smallskip

\hrule\smallskip
\noindent
Part II of two parts to appear in J. Stat. Phys. in honor of Elliott H. Lieb's 70th birthday;
received by JSP on Nov.18, 2003 
accepted on March 05, 2004.\\
\noindent
\copyright{2004} 
The author. This paper may be reproduced for noncommercial purposes.

\newpage

	\section{Introduction}

 In our previous paper
   \cite{KiePapI}
we presented the first relativistic classical electromagnetic field theory
in which the notion of the (spin-less) point electron is satisfactorily implemented.
  The classical theory formulated in
   \cite{KiePapI}
is divergence problem-free, a-priori speaking; in particular, no regularization 
or renormalization is needed to give sense to the basic variables of the theory.
 Half of this feat was actually accomplished long ago,\footnote{Our paper
                          \cite{KiePapI}
			  contains a fairly exhaustive collection of scientific and
			  bibliographical background information which is also 
			  pertinent to the present paper.}
by Born and Infeld
       \cite{BornInfeldB}.
 Their nonlinear Maxwell--Born--Infeld field equations eliminate the infinite 
classical electromagnetic self-energy problem for point charges, and since  
  \cite{Boillat, Plebanski}
we know that they do so in a compellingly unique way.
 Unfortunately, although no longer diverging, the Lorentz self-force remained ill-defined 
in magnitude and direction. 
 It is not the case, as the founding fathers of that theory believed, that 
this Lorentz self-force problem could be overcome by simply regularizing  and then 
taking limits, and / or imposing energy conservation.
 Thus, contrary to Dirac's early proclamation that 
``[t]he classical theory is found to be completely satisfactory''
  (\cite{diracBI}, p.32),
a satisfactory law of motion for the point charges has in fact been missing.
 Our main contribution in 
          \cite{KiePapI}
is to supply a well-defined law of point charge motion.
 A relativistic many-body Hamilton--Jacobi equation has to be solved together with a 
system of Maxwell--Born--Infeld field equations for generic point charge sources; i.e. 
instead of the actual Maxwell--Born--Infeld fields one studies a whole family of
such fields indexed by their generic point sources configuration. 
 The Hamilton--Jacobi guiding law is solved subsequently to get the actual 
particles' motions, and when this actual point charges configuration is substituted
for the generic configuration in the indexed fields, the actual electromagnetic
Maxwell--Born--Infeld fields are obtained. 

 This classical electromagnetic theory with point electrons is in itself 
an interesting object for further study, but the really interesting question 
is whether it can serve as stepping stone en route to a consistent
quantum theory of electromagnetism with point electrons.
 For Born and Infeld
     \cite{BornA, BornInfeldA, BornInfeldB, BornB, BornInfeldC,  BornC},
Pryce 
                \cite{PryceB, PryceC, PryceD, PryceF},
Schr\"odinger
		\cite{ErwinEiHBiD, ErwinDUBLINa, ErwinDUBLINb, ErwinDUBLINc},
and Dirac
                \cite{diracBI},
this was the driving force behind their quest, but the attempts in 
   \cite{BornInfeldC, PryceF, diracBI}
to quantize the Maxwell--Born--Infeld field equations revealed that
``difficulties arise with the passage to the quantum theory, 
which appear to be insoluble with present methods of quantization'' 
  (\cite{diracBI}, p.32),
and 
``[t]he adaption of these ideas to the principles of quantum theory and the 
introduction of the spin has [...] met with no success'' 
           (\cite{BornD}, p.375).
 One reason for the failure of these attempts is of course the fact that the 
Maxwell--Born--Infeld field equations with point charges do not in themselves constitute 
a complete classical dynamical theory, but this is not the only reason. 
 The other, not less important reason is their choice of quantization procedure, which was
patterned after the available standard procedure of replacing classical quantities by operators.
 However, if one wants to use a classical divergence problem-free electromagnetic field 
theory with point electrons as point of departure for the construction of a divergence 
problem-free electromagnetic quantum theory with point electrons, 
then one can reasonably hope to be successful only if one tries \textit{ not to tamper} 
with the integrity of the mathematical structures which are responsible for the absence 
of any divergence problems at the classical level. 
 In this spirit we have applied a `least invasive quantization procedure' to
the classical electromagnetic field theory developed in our previous paper
   \cite{KiePapI}. 
 The electromagnetic quantum theory with point electrons which results from this
is the subject of the present paper.

  Like the classical theory from which it springs, the quantum theory 
describes the joint dynamics of spin-less point charges and the total electromagnetic fields.
 The point charges move according to a relativistic generalization of the
first order guiding equation alluded to by Born\footnote{Born  
                                            seems to have favored a stochastic guiding equation but remarked
					       that Frenkel had pointed out the possibility
					       of a deterministic guiding equation.}
in 
                                                \cite{BornsQMpapers}
and the explicit form of which de Broglie
              \cite{deBroglieB}
and Bohm 
             \cite{BohmsHIDDENvarPAPERS}
discovered subsequently. 
 The guiding field is the gauge-invariant gradient of the phase of 
a wave function solving a relativistic Klein--Gordon equation for the electromagnetic potentials
of the total electromagnetic fields indexed by generic point sources, which in turn are obtained from the 
Maxwell--Born--Infeld field equations with generic point charges as sources.
 The guiding equation is to be solved subsequently to obtain the motion of the actual point
charges, and when this actual point charges configuration is substituted for the generic
ones in the indexed fields, the actual electromagnetic Maxwell--Born--Infeld fields are obtained.
  While this \emph{partially} quantized theory is certainly only a modest step forward,
for spin and photon are not yet incorporated, the theory is a priori 
free of any divergence problems; hence, once again there is no need 
for regularization or, for that matter, renormalization. 
 We take this as a major encouragement to pursue the full quantization, with 
spin and photon, in due course. 

  In this paper we also re-address the subtle issue of the value of  
`Born's aether constant,' the new dimensionless physical constant that 
enters the Born--Infeld law of the `aether.'\footnote{In our previous paper 
                                                      \cite{KiePapI} 
						      we stipulated that `aether' is short for 
						      `electromagnetic vacuum.'}
 In 
 \cite{KiePapI} 
we found that Born's reasoning
      \cite{BornA}
that the value of this aether constant be chosen so that
the empirical electron rest energy $\me c^2$ equals the now finite 
electrostatic energy of a point charge at rest, 
is not conclusive at the classical level. 
 In principle the value of the aether constant $\beta$ should be inferable from 
the spectral data, but that means its true value will be computable only
after spin, and perhaps even the photon, are implemented into the theory.
 Nevertheless, by discussing the `spin-less hydrogen' spectrum in some 
detail we here find some decent upper bounds on Born's aether constant 
that, curiously, still leave the value computed in 
              \cite{BornA, BornInfeldA} 
viable, for now.
 Incidentally, for our discussion of the hydrogen spectrum we also prove the
first rigorous two-body results for the nonlinear Maxwell--Born--Infeld field equations.

  In the remainder of this paper, we first present the least invasive
quantization of the classical theory, using the compact, 
manifestly Poincar\'e- and Weyl-covariant formalism.
  We will then discuss the spin-less hydrogen spectrum, for which purpose
we prove the first rigorous results  for the classical 
Maxwell--Born--Infeld field equations with two point charges.
    The paper concludes, after a summary, with an outlook and an epilogue 
in celebration of Elliott H. Lieb's $70^{\mathrm{th}}$ birthday.

\bigskip
        \section{The electromagnetic quantum theory in covariant format}

                \subsection{The basic equations}

  As in 
   \cite{KiePapI}
we use dimensionless units with the following conversion factors between Gaussian 
and dimensionless units: $\hbar$ (Planck's constant divided by $2{\pi}$) 
for both the unit of action and the magnitude of angular momentum, 
$e$ (elementary charge) for the unit of charge, 
$\me$ (electron rest mass) for the unit of mass, 
$c$ (speed of light \textit{in vacuo}) for the unit of speed.
  Thus, length and time are both referred to in the same dimensionless unit, 
multiples of the Compton wave length of the electron $\lC = {\hbar}/{\me}c$.
  Accordingly, the unit magnitude of the electromagnetic 
fields is to be converted by a factor $e/\lC^2$, 
while the natural unit for the magnitude  of momentum and the 
energy are converted, respectively, by factors ${\me}c$ and ${\me}c^2$.
 The parameter $\alpha$ will denote  Sommerfeld's fine structure constant.

      \subsubsection{The equations of the flat electromagnetic spacetime}

 In our partially quantized theory, the  (flat) \textit{electromagnetic spacetime} 
structure is defined as in our classical theory.
 Thus, Minkowski spacetime $\MM^4$ is made into an electromagnetic spacetime
by decorating it with a classical electromagnetic field which satisfies the 
Maxwell--Born--Infeld field equations in a distributional sense; the 
field may not be well defined along one-dimensional time-like defects. 
 When cut with a space-like slice, the electromagnetic field is finite 
in a punctured space-like neighborhood of these line defects, which 
themselves are noticeable as moving point charges in the space-like slice(s).
 To have the quantum theory minimally self-contained, we briefly recall these 
laws in the genuinely electromagnetic setting, in which all 
point charges are positive or negative unit charges, representing 
electrons of either variety.

 Let $\Eta_k$ be the point history (future oriented time-like world-line)
of the $k$-th particle, and let $\bigcup_k\Eta_k$ denote the set of $N$
point histories with which $\MM^4$ is threaded. 
 Faraday's electromagnetic field tensor $\FQ$ is a two-form on 
$\MM^4\backslash\bigcup_k\Eta_k$ satisfying the \emph{Faraday--Maxwell law} 
                 \cite{WheeleretalBOOK}
$
        \dQ{\FQ} 
=
\MTWvec{0}
$
in the sense of distributions.
 Let $\Hodge\FQ$ (etc.) be the Hodge dual of $\FQ$ (etc.).
 Then the \emph{Born and Infeld law of the aether}, 
\beq
-\Hodge\MQ 
= 
\frac{\FQ - \beta^4  {}\Hodge\big(\FQ\wedge\FQ \big){}\Hodge\FQ} 
     {\sqrt{1 - \beta^4 {}\Hodge\big(\FQ\wedge \Hodge\FQ \big)
              - \beta^8 \left({}\Hodge \big(\FQ\wedge\FQ \big)\right)^2 }}
\,,
\label{eq:GEOaetherLAWborninfeld}
\eeq
in which $\beta\in(0,\infty)$ is Born's aether constant,
maps $\FQ$ to Maxwell's electromagnetic displacement tensor $\MQ$, 
which is a two-form on $\MM^4\backslash\bigcup_k\Eta_k$ satisfying the 
\emph{Amp\'ere--Coulomb--Maxwell} law
                 \cite{WheeleretalBOOK}
$ 
\dQ\MQ 
= 
4\pi \JQ 
$
in the sense of distributions, where (cf. 
                 \cite{jacksonBOOK, WheeleretalBOOK, ThirringBOOKa})
\begin{equation}
         \JQ(\varpi) 
=
	\sum_{k\in\cN} \int_{-\infty}^{+\infty}
		\pmk\Hodge\uQ_k(\tau)\delta_{\eta_k(\tau)}\big(\varpi\big)\,\dd\tau
\, ,
\label{eq:GEOptchargecurrent}
\end{equation}
is  the electromagnetic current density at $\varpi\in\MM^4$ 
of a system of $N\geq 0$ electric unit point charges.
    Here, $\delta_{\eta(\tau)}\big(\,.\,\big)$ 
is the Dirac measure on $\MM^4$ concentrated at $\eta(\tau)$, 
where $\tau$ is a Lorentz-scalar time parameter. 
 If $\tau$ is proper-time, then  $\uQ_k(\tau)$ is the 
future-oriented Minkowski-velocity co-vector, which is 
the metrical dual of the Minkowski-velocity vector 
$\dd\varpi/\dd\tau|_{\varpi=\eta_k(\tau)}$
of the $k^{\mathrm{th}}$ point charge, and 
$\Hodge\uQ_k(\tau)$ is the Hodge dual of  $\uQ_k(\tau)$.
 Furthermore, $\pmk$ is the sign of that charge.
 Also, $\cN\subset\NN\cup\{0\}$ is the set of $N$ indices, and
we set $\cN\equiv\emptyset$ if $N=0$, in which case
$\sum_{k\in\emptyset} (...) \equiv 0$, so
that the charge-free situation is included in \refeq{eq:GEOptchargecurrent}.
 It is well-known, and readily verified, that  \refeq{eq:GEOptchargecurrent}
satisfies the \emph{law of the conservation of electric charge},
$ 
  \dQ\JQ 
=
  \MTWvec{0}
$ 
in the sense of distributions, as demanded by the 
Amp\'ere--Coulomb--Maxwell law.

 By the manifestly covariant character of the Maxwell--Born--Infeld field laws with point
sources, all Lorentz observers of any particular, actually realized electromagnetic structure 
in Minkowski spacetime satisfying these laws would necessarily conclude that they see their 
respective Lorentz frame manifestations of the \emph{same} electromagnetic spacetime, whatever 
their relative states of uniform motion with respect to each other might be.

 We end this subsection by recalling that $\FQ$ is also exact and can therefore be written as 
the exterior derivative of a one-form, i.e. $\FQ=\dQ\AQ$,
where $\AQ$ is the \emph{electromagnetic potential} on ${\MM}^{4}$; notice that
while the exterior derivative is to be understood in the sense of distributions, away from the
location of a charge the derivative exists in the regular sense, and since the singularities
of $\FQ$ are mild discontinuities, the one-form $\AQ$ can be extended continuously into the
locations of the point charges.
 With the help of $\AQ$ the law of motion for the point charges can be formulated.

               \subsubsection{The law of motion}

 The partial quantization of the classical theory is achieved by modifying the
\emph{law of motion} for the point charges.
 Like its classical counterpart, the quantum law of motion is formulated 
on $\MM^{4N}_{\neq}$, the configuration space of $N$ ordered 
world-points $\MM^{4N} =\times_{k=1}^N \MM^4_k$ with all co-incidence 
points removed; here $\MM^4_k$ is the $k^{\mathrm{th}}$ copy of $\MM^4$.
 We first recall these basic ingredients of the classical law of motion.
 Its reformulation in terms of bundles on $\MM^{4N}_{\neq}$ will then
naturally suggest a generalization which will reveal itself as a legitimate
quantum law of motion for a single relativistic particle, and 
in the formal non-relativistic particle limit for many particles; 
modifications are necessary, however, for many relativistic 
particles. 

\medskip
\textit{The classical law of motion and its differential-geometrical reformulation}

 In 
 \cite{KiePapI}
we postulated fields $^\sharp\AQ$ on $\MM^4\times\MM^{4N}_{\neq}$
such that $^\sharp\AQ(\varpi,\varpi_1,...,\varpi_N)$
reduces to $\AQ(\varpi)$ when the $N$ configuration world-points 
$\varpi_k\in\Eta_k$, $k=1,...,N$; such fields can be defined only 
w.r.t. some space-like foliation of $\MM^4$, by which we mean that
$^\sharp\AQ(\varpi,\varpi_1,...,\varpi_N)$ reduces to 
$\AQ(\varpi)$ when in addition to $\varpi_k\in\Eta_k$, $k=1,...,N$, 
the $N+1$ world-points $\{\varpi,\varpi_1,...,\varpi_N\}$ are picked 
from a constant-time $t$ leaf of the smooth foliation; eventually we 
worked with the standard foliation.
 The subset in $\MM^{4N}_{\neq}$ of $t$ leaf-synchronized configurations, 
which is diffeomorphic to $\RR \times \RR^{3N}_{\neq}$,
will be called $t$-synchronized $\MM^{4N}_{\neq}$, for brevity.
 Similarly we defined fields $^\sharp\FQ = \dQ {\,}{^\sharp\!\AQ}$, etc. 
 This generic world configuration-indexed family of electromagnetic potentials
$^\sharp\AQ$  defines $N$ fields 
$\widetilde\AQ_k(\varpi_1,...,\varpi_N) \equiv 
{\,}{^\sharp\!\AQ}(\varpi_k,\varpi_1,...,\varpi_N)$, $k=1,...,N$.
 We further postulated that there is a scalar field $\widetilde\Phi$
on $\MM^{4N}_{\neq}$ such that 
${\dQ}_k \widetilde\Phi(\varpi_1,...,\varpi_N)
-
\pmk {\alpha}\widetilde\AQ_k(\varpi_1,...,\varpi_N)$
is a Minkowski velocity co-vector field when the $N$ world-points $\varpi_1,...,\varpi_N$ 
are varied over a leaf of the foliation associated with the $^\sharp$fields. 
 It follows that $\widetilde\Phi$ has to obey the $N$ equations
\beq
\Hodge
\Bigl( \bigl( 
{\dQ}_k \widetilde\Phi
-\pmk {\alpha}\widetilde\AQ_k
\bigr)\wedge
\Hodge \bigl( 
{\dQ}_k \widetilde\Phi
-\pmk {\alpha}\widetilde\AQ_k
\bigr)\Bigr)
 = 1,
\label{eq:GEOhamjacPDE}
\eeq
understood w.r.t. the foliation. 
 Equations \refeq{eq:GEOhamjacPDE}, which have a double root of which the future-oriented 
one is to be chosen, determine how the Minkowski-velocity co-vector fields evolve from one 
leaf of the foliation to another.
 The Minkowski-velocities co-vector field yields a first-order guiding law  for the
world-points $\varpi_k = \eta_k(\tau)$ at proper-time $\tau$; namely
their Minkowski velocity co-vectors $\uQ_k(\tau) = \dd\varpi_k/\dd\tau\alongHk$ 
obey
\beq
\uQ_k(\tau)
=
{\dQ}_k \widetilde\Phi(\varpi_1,...,\varpi_N)
-
\pmk {\alpha}\widetilde\AQ_k(\varpi_1,...,\varpi_N)
\,,
\label{eq:GEOhamjacLAW}
\eeq
where the $\varpi_1,...,\varpi_N$ here are the world-points of the point charges on the same leaf. 

 While the $^\sharp$fields and $\widetilde\Phi$ are defined w.r.t. a foliation,
as explained at the beginning of the first subsection of 2.1.2,
the actual electromagnetic output of the theory is Poincar\'e co-variant, as 
mentioned in the previous subsection. 
 The theory is also manifestly Weyl-covariant, i.e. a gauge transformation
\bea
^\sharp\AQ(\varpi,\varpi_1,...,\varpi_N)  
\!\!\!
&\to&
\!\!\!
^\sharp\AQ(\varpi,\varpi_1,...,\varpi_N) +\dQ\Upsilon(\varpi) 
\label{eq:GEOgaugetrA}
\\
	\widetilde\Phi(\varpi_1,...,\varpi_N) 
\!\!\!
&\to&
\!\!\!
	\widetilde\Phi(\varpi_1,...,\varpi_N) +  \sum_k \pmk \alpha\Upsilon(\varpi_k) 
\,,
\label{eq:GEOgaugetrPHI}
\eea
with any zero-form $\Upsilon:\MM^4\to\RR$ leaves $\FQ$ and the $\uQ_k$ invariant.

 Next we recast \refeq{eq:GEOhamjacLAW} into a more geometrical format. 
 Let $\UQ = (\uQ_1,...,\uQ_N)$ consist of the $N$ Minkowski-velocity co-vectors 
$\uQ_k$ at the $N$ world-points $\varpi_k$ of the point-histories $\Eta_k$, $k=1,...,N$, 
piercing a leaf.
 Furthermore, set $\widetilde\AQ= (\widetilde\AQ_1,...,\widetilde\AQ_N)$, and let 
$Z$ be a diagonal matrix, the $k^{\mathrm{th}}$ $4\times 4$ block of which having entries 
$z_k$ (in its diagonal), and let $\dQ_{(N)}$ be Cartan's exterior derivative on $\MM^{4N}$.
 With these definitions, it is now readily seen that the first order guiding laws 
\refeq{eq:GEOhamjacLAW} can be restated as
\beq
\UQ
=
e^{-i\widetilde\Phi}
  \left( -i\dQ_{(N)} -{\alpha}Z\widetilde\AQ \right)
e^{i\widetilde\Phi}
\,,
\label{eq:GEOhamjacLAWconfigspaceUNITARY}
\eeq
i.e. the actual $\UQ$ is set equal to the covariant logarithmic derivative, at the
actual world-configuration on a leaf of a foliation, of a unitary section $e^{i\widetilde\Phi}$ 
of a complex line bundle on foliation-synchronized
$\MM^{4N}_{\neq}$, with $({\alpha}Z)\widetilde\AQ$ the \textit{electromagnetic connection} 
on the bundle.\footnote{The $k^{\mathrm{th}}$ block component of the 
                        \emph{electromagnetic curvature} of the bundle is 
			(upto a factor $\pmk\alpha$)
			$\dQ_k\widetilde\AQ_k =\widetilde\FQ_k$. 
			Note that $\widetilde\FQ_k(\varpi_k)\neq \FQ(\varpi_k)$, 
			for the r.h.s. in this non-equation is ill-defined.}

\medskip
\textit{The quantum law of motion}

  To begin with, we remark that the requirement that the section of the complex line 
bundle be unitary can be dropped without changing the content of the classical theory. 
  Thus, instead of $e^{i\widetilde\Phi}$ we may consider a complex 
$\widetilde\psi \equiv \widetilde\varrho^{\,1/2}e^{i\widetilde\Phi}$, where
$\widetilde\varrho^{\,1/2}$ 
is a positive amplitude function on $\MM^{4N}_{\neq}$, 
in terms of which \refeq{eq:GEOhamjacLAWconfigspaceUNITARY} becomes
\beq
\UQ
=
\Re\left(
\widetilde\psi^{\,-1}
  \left( -i\dQ_{(N)} -{\alpha}Z\widetilde\AQ \right)\widetilde\psi\right)
\,,
\label{eq:GEOhamjacLAWconfigspace}
\eeq
where $\Re$ means `real part.'
  Superficially \refeq{eq:GEOhamjacLAWconfigspace} has a formal 
de Broglie--Bohm like appearance, but appearances are misleading, for so far
$\widetilde\varrho^{\,1/2}$  remains an undetermined, superfluous
mathematical luxury which has no input into the theory whatever. 
 Indeed, the condition that the $N$ Minkowski co-vector components of 
the right side of \refeq{eq:GEOhamjacLAWconfigspace} give 
future-oriented time-like unit co-vector fields just gives back the Hamilton--Jacobi 
law for the phase $\widetilde\Phi$ of the bundle section, and
\refeq{eq:GEOhamjacLAWconfigspace} generates just the classical motions.

 Next, we remark that we can re-calibrate the particles' time parameter $\tau$ in 
\refeq{eq:GEOptchargecurrent} from being proper time into any other Lorentz-scalar time
without changing the actual electromagnetic spacetime structure of the theory, whether
it comes from a classical motion or not. 
 To remain at the classical level of electromagnetic theory, the $N$ Minkowski 
co-vector components of the right side of \refeq{eq:GEOhamjacLAWconfigspace} 
then have to produce future-oriented time-like co-vector fields which are
compatible with the new normalization of $\UQ$ which is brought about 
by the re-calibration of $\tau$ in the classical motions; of course, the 
$\uQ_k$ are now in general no longer unit co-vectors. 
 Moreover, the re-calibration will also entail a change of the foliation 
w.r.t. which the law of motion in configuration space is constructed. 
 All this will change the appearance of the equations for $\widetilde\Phi$, 
and the appearance of the law of motion, but it will not change the classical 
content of the theory.

 Things will change, however, if we proceed with a re-calibration of $\tau$ and the 
corresponding adjustments in the normalization of the guiding law on $\MM^{4N}_{\neq}$
without adjusting the foliation. 
 In particular, since nothing prevents us a-priori from linking the re-calibration 
of $\tau$ to the amplitude field $\widetilde\varrho^{\,1/2}$ on $\MM^{4N}_{\neq}$, 
now contemplate the following $\widetilde\varrho^{\,1/2}$-dependent 
normalization of the future-oriented time-like co-vector fields defined by the
r.h.s. of \refeq{eq:GEOhamjacLAWconfigspace}, namely for the $k$th component
\beq
\Hodge
\Bigl( \bigl( 
{\dQ}_k \widetilde\Phi
-\pmk {\alpha}\widetilde\AQ_k
\bigr)\wedge
\Hodge \bigl( 
{\dQ}_k \widetilde\Phi
-\pmk {\alpha}\widetilde\AQ_k
\bigr)\Bigr)
 = 1 - \widetilde\varrho^{\,-1/2}{\mathbf{\Box}}_k \widetilde\varrho^{\,1/2},
\label{eq:GEOhamjacPDErecalib}
\eeq
where ${\mathbf{\Box}}_k$ is the wave operator for the $k$th world-point variable.
 The r.h.s. of \refeq{eq:GEOhamjacPDErecalib} evaluated a-posteriori for the actual 
motions then yields the re-calibration of $\tau$ along $\Eta_k$ via
$\Hodge(\uQ_k\wedge\Hodge\uQ_k) 
= 1 - \widetilde\varrho^{\,-1/2}{\mathbf{\Box}}_k \widetilde\varrho^{\,1/2}\alongHk$.
  At this point, the amplitude field $\widetilde\varrho^{\,1/2}\geq 0$ may still be 
just arbitrarily prescribed.
 But now assume that instead of prescribing $\widetilde\varrho$, we also postulate
$N$ individual `continuity equations' for $\widetilde\varrho$, namely
\beq
\dQ_k
\Bigl(
\widetilde\varrho\;\; {\Hodge}\!\!\left({\dQ}_k \widetilde\Phi
-
\pmk {\alpha}\widetilde\AQ_k\right) \Bigr) 
= 
0\,,
\label{eq:kTHelectronContinuityEQrecalib}
\eeq
of course all of them understood w.r.t. a foliation. 
 All these seem to be quite innocuous changes that just implement the freedom of having an 
amplitude factor $\widetilde\varrho^{\,1/2}$ and a re-calibration of $\tau$ `at our disposal.' 
 And yet, considered first for a single electron, and w.r.t. the standard foliation of spacetime, 
this actually achieves a quantization of the classical theory, albeit only a partial one, for
spin and photon are not yet incorporated. 
 Indeed, with respect to the standard foliation of spacetime, 
the section of the complex line bundle now satisfies a Klein--Gordon equation, 
which is the `correct' relativistic wave equation for a spin-less point electron 
in interaction with electromagnetic fields.
  The quotes around `correct' are meant to remind the reader that there are no spin-less
electrons in Nature, for which reason we do not really know what the
correct wave equation would be if there were such a beast. 
  Moreover, and the Pauli--Weisskopf work on the quantum field-theoretical
interpretation of the Klein--Gordon equation aside, it is well-known that the 
quantum mechanical interpretation of the Klein--Gordon equation is burdened 
with the problem that this equation is of second order in time, not of first order.
 Be that as it may, it is certainly not an unreasonable first step to verify that
the electromagnetic Maxwell--Born--Infeld field equations with point charge source
can be consistently coupled with a de Broglie--Bohm like law of motion generated by
a Klein--Gordon equation.

 Unfortunately, the straightforward extension to the many electrons case, as formally depicted 
in \refeq{eq:GEOhamjacPDErecalib} to \refeq{eq:kTHelectronContinuityEQrecalib}, reveals 
problems with the synchronization that require the input of new ideas which lead
to modifications of the simple scheme laid down above, and on which we briefly 
comment later on in the paper. 
 Rewardingly though, if taken as heuristic starting point for a non-relativistic 
approximation to the equations of motion, formulas 
\refeq{eq:GEOhamjacPDErecalib} to \refeq{eq:kTHelectronContinuityEQrecalib}
are quite useful and lead to the many body Schr\"odinger equation coupled with 
the Maxwell--Born--Infeld $^\sharp$field equations.

\section{$\!\!\!$The~$\!$electromagnetic$\!$~quantum$\!$~theory$\!$~on$\!$~a$\!$~standard$\!$~foliation}

	\subsection{The actual electromagnetic field laws}

 The field equations for the actual electromagnetic fields on spacetime $\MM^4$
are the Maxwell--Born--Infeld equations as listed in Sect. 4.2 of 
          \cite{KiePapI}.
 Very briefly, given point charges source terms
\bea
        j (t,{\SPvec{s}})
\!\!\!&=&\!\!\!
{\textstyle\sum_{k\in\cN}} \pmk\delta_{{\SPvec{r}}_{k}(t)}({\SPvec{s}}),
\label{eq:FOLIpointSOURCESrho}
\\
        {\SPvec{j}}(t,{\SPvec{s}})
\!\!\!&=&\!\!\!
{\textstyle\sum_{k\in\cN}}
\pmk\delta_{{\SPvec{r}}_{k}(t)}({\SPvec{s}})\bulldif{\SPvec{r}}{_k}(t)
\,,
\label{eq:FOLIpointSOURCESjjj}
\eea
the evolution equation for ${\SPvec{D}}$,
\beq
        {\partial}{\SPvec{D}}
=
        {\nabla}\times{\SPvec{H}}
-
	 4\pi         {\SPvec{j}}
\,,
\label{eq:FOLIeD}
\eeq
constrained by 
\beq
        {\nabla}\cdot {\SPvec{D}}
=
         4 \pi         j 
\,,
\label{eq:FOLIcD}
\eeq
has to be solved together with the evolution equation for  ${\SPvec{B}}$,
\beq
\partial \SPvec{B} = -\nabla\times\SPvec{E}
\label{eq:FOLIeB},
\eeq
constrained by
\beq
\nabla\cdot\SPvec{B} =0
\label{eq:FOLIcB}, 
\eeq
where the fields ${\SPvec{E}}$ and ${\SPvec{H}}$ are defined in terms of
${\SPvec{B}}$ and ${\SPvec{D}}$ by the aether laws 
\bea
&&
{\SPvec{E}} 
= 
\frac{
{\SPvec{D}}-\beta^4{\SPvec{B}}\times({\SPvec{B}}\times{\SPvec{D}})
}{
\sqrt{    1 
	+ \beta^4(|{\SPvec{B}}|^2 + |{\SPvec{D}}|^2) 
	+ \beta^8|{\SPvec{B}}\times {\SPvec{D}}|^2
      }}
\label{eq:FOLIeqEofBD}
\\
&&
{\SPvec{H}} 
 = 
\frac{
{\SPvec{B}}-\beta^4{\SPvec{D}}\times({\SPvec{D}}\times {\SPvec{B}})
     }{
\sqrt{	  1 
	+ \beta^4(|{\SPvec{B}}|^2 + |{\SPvec{D}}|^2) 
	+ \beta^8|{\SPvec{B}}\times {\SPvec{D}}|^2
      }}
\label{eq:FOLIeqHofBD}
\, 
\eea
of Born and Infeld 
		\cite{BornInfeldB},
with $\beta\in (0,\infty)$. 
 
 Once ${\SPvec{B}}$ and ${\SPvec{E}}$ are known, one can also compute a 
magnetic vector potential $\SPvec{A}$ satisfying the evolution equation 
\beq
\partial \SPvec{A} = -\nabla A - \SPvec{E}
\label{eq:FOLIeAmagn}, 
\eeq
and the constraint equation 
\beq
\nabla\times\SPvec{A} = \SPvec{B}
\label{eq:FOLIcAmagn},
\eeq
where $A$ satisfies the evolution equation of the Lorentz--Lorenz gauge, 
\beq
\partial {A} = -\nabla\cdot\SPvec{A}
\label{eq:FOLIeAelec}.
\eeq

 We recall that in the absence of any source terms, the
resulting charge-free Maxwell--Born--Infeld field equations
form a closed system of equations for the actual electromagnetic
fields in $\MM^4$. 
 Also, the Maxwell--Born--Infeld field equations with \emph{given}
point charge sources are well-posed locally 
and can be solved `bottom-up,' at least in principle. 
 Of course, the actual point charge sources in $\MM^4$ are only known 
after the motions of all point charges have been computed. 
 As emphasized in the covariant section, this requires solving a 
whole family of Maxwell--Born--Infeld $^\sharp$field equations with
generic point sources. 
 The solutions for the $^\sharp{\SPvec{A}}$ and $^\sharp{A}$ 
then define the solutions of
\refeq{eq:FOLIeAmagn} and \refeq{eq:FOLIeAelec}
through conditioning with the actual motion, so that 
the actual Maxwell--Born--Infeld field equations get 
solved `top-down,' then. 
 
	\subsection{The $t$-synchronized $^\sharp$fields equations}
 
 Also the $^\sharp$fields equations are unaltered; cf. 
 \cite{KiePapI}.
 The  $t$-synchronized space and time decomposition of 
$^\sharp\AQ(\varpi,\varpi_1,...,\varpi_N)$ into components 
$^\sharp{A}$ and $^\sharp{\SPvec{A}}$ gives 
\beq
{{A}}^\sharp(t,\SPvec{s},\SPvec{S})
\equiv
{}^\sharp{{A}}(t,\SPvec{s}, t_1,\SPvec{s}_1,...,t_k,\SPvec{s}_k,...,t_N,\SPvec{s}_N)
\big|_{t_1=t_2=...=t_N=t}\big.,
\eeq
\beq
{\SPvec{A}}^\sharp(t,\SPvec{s},\SPvec{S})
\equiv
{}^\sharp{\SPvec{A}}(t,\SPvec{s}, t_1,\SPvec{s}_1,...,t_k,\SPvec{s}_k,...,t_N,\SPvec{s}_N)
\big|_{t_1=t_2=...=t_N=t}\big.
\eeq
on $\RR\times\RR^{3(N+1)}$ (etc. for the other $^\sharp$fields).
 As stipulated earlier, by conditioning with the actual configuration we 
want to obtain the actual fields on $\MM^4$ (in Lorentz gauge, say), 
i.e. ${A}^\sharp (t,\SPvec{s},\SPvec{R}(t)) = A(t,\SPvec{s})$ and 
${\SPvec{A}}^\sharp (t,\SPvec{s},\SPvec{R}(t))= {\SPvec{A}}(t,\SPvec{s})$ (etc.).
 This canonically fixes the equations for the $t$-synchronized space and time 
decomposition of the $^\sharp$fields. 
 Namely, ${{A}}^\sharp(t,\SPvec{s},\SPvec{S})$, ${\SPvec{A}}^\sharp(t,\SPvec{s},\SPvec{S})$, 
and ${\SPvec{D}}^\sharp(t,\SPvec{s},\SPvec{S})$ satisfy the evolution equations
\beq
\partial{{A}}^\sharp(t,\SPvec{s},\SPvec{S})
=
- \SPvec{V}(t,\SPvec{S})\cdot \nabla_{\!\!\SPvec{S}} 
{A}^\sharp(t,\SPvec{s},\SPvec{S})
- \nabla\cdot \SPvec{A}^\sharp(t,\SPvec{s},\SPvec{S}) ,
\label{eq:FOLIeqAscalSHARP}
\eeq
\beq
\partial {\SPvec{A}}^\sharp(t,\SPvec{s},\SPvec{S})
=
- \SPvec{V}(t,\SPvec{S})\cdot\nabla_{\!\!\SPvec{S}} 
 {\SPvec{A}}^\sharp(t,\SPvec{s},\SPvec{S})
- \nabla A^\sharp(t,\SPvec{s},\SPvec{S}) -  {\SPvec{E}}^\sharp(t,\SPvec{s},\SPvec{S}),
\label{eq:FOLIeqAvectSHARP}
\eeq
\beq
\partial {\SPvec{D}}^\sharp(t,\SPvec{s},\SPvec{S})
=
- \SPvec{V}(t,\SPvec{S})\cdot\nabla_{\!\!\SPvec{S}} 
 {\SPvec{D}}^\sharp(t,\SPvec{s},\SPvec{S})
+ \nabla \times {\SPvec{H}}^\sharp(t,\SPvec{s},\SPvec{S}) 
- 4\pi {\SPvec{j}}^\sharp(t,{\SPvec{s}},{\SPvec{S}}),
\label{eq:FOLIeqDvectSHARP}
\eeq
where 
$\SPvec{V}(t,\SPvec{S})\cdot\nabla_{\!\!\SPvec{S}} \equiv \sum_{k=1}^N \SPvec{v}_k(t,\SPvec{S})\cdot\nabla_k$
is the velocity field on configuration space that still needs to be defined;
furthermore, ${\SPvec{D}}^\sharp(t,\SPvec{s},\SPvec{S})$ obeys the constraint equation
\beq
\nabla\cdot {\SPvec{D}}^\sharp(t,\SPvec{s},\SPvec{S})
=
 4\pi {{j}}^\sharp(t,{\SPvec{s}},{\SPvec{S}}),
\label{eq:FOLIeqDvectSHARPconstraint}
\eeq
where\footnote{Recall that the $^\sharp$field re-formulation of the continuity equation 
	       of the charge conservation (in spacetime), 
	       $\partial{{j}}^\sharp(t,\SPvec{s},\SPvec{S})
	       = - \SPvec{V}(t,\SPvec{S})\cdot \nabla_{\!\!\SPvec{S}} 
	       {j}^\sharp(t,\SPvec{s},\SPvec{S})
	       - \nabla\cdot \SPvec{j}^\sharp(t,\SPvec{s},\SPvec{S})$, 
	       is an identity, not an independent equation.}
\bea
        j^\sharp (t,{\SPvec{s}},{\SPvec{S}})
\!\!\!&=&\!\!\!
{\textstyle\sum_{k\in\cN}} \pmk\delta_{{\SPvec{s}}_{k}}({\SPvec{s}}),
\label{eq:FOLIpointSOURCESrhoSHARP}
\\
&&
\nonumber
\\
        {\SPvec{j}}^\sharp(t,{\SPvec{s}},{\SPvec{S}})
\!\!\!&=&\!\!\!
{\textstyle\sum_{k\in\cN}}
\pmk\delta_{{\SPvec{s}}_{k}}({\SPvec{s}}){\SPvec{v}}{_k}(t,{\SPvec{S}})
\, .
\label{eq:FOLIpointSOURCESjjjSHARP}
\eea
 The fields 
${\SPvec{E}}^\sharp(t,\SPvec{s},\SPvec{S})$ and 
${\SPvec{H}}^\sharp(t,\SPvec{s},\SPvec{S})$ in 
\refeq{eq:FOLIeqAvectSHARP}, \refeq{eq:FOLIeqDvectSHARP}
are defined in terms of 
${\SPvec{D}}^\sharp(t,\SPvec{s},\SPvec{S})$ 
and 
${\SPvec{B}}^\sharp(t,\SPvec{s},\SPvec{S})$ 
in precisely the same manner as the actual fields 
${\SPvec{E}}(t,\SPvec{s})$ and ${\SPvec{H}}(t,\SPvec{s})$ 
are defined in terms of 
${\SPvec{D}}(t,\SPvec{s})$ and ${\SPvec{B}}(t,\SPvec{s})$ 
through the Born--Infeld aether laws 
\refeq{eq:FOLIeqEofBD}, \refeq{eq:FOLIeqHofBD}, 
while 
${\SPvec{B}}^\sharp(t,\SPvec{s},\SPvec{S})$ 
in turn is defined in terms of 
${\SPvec{A}}^\sharp(t,\SPvec{s},\SPvec{S})$ 
in precisely the same manner as the actual 
${\SPvec{B}}(t,\SPvec{s})$ 
is defined in terms of the actual
${\SPvec{A}}(t,\SPvec{s})$ 
in 
\refeq{eq:FOLIcAmagn}.

 It is straightforward to verify that by substituting the actual configuration 
$\SPvec{R}(t)$ for the generic $\SPvec{S}$ in the $t$-synchronized $^\sharp$fields 
satisfying the above equations, we obtain the actual electromagnetic potentials, 
fields, and charge-current densities satisfying the Maxwell--Born--Infeld field 
equations (in Lorentz--Lorenz gauge). 

	\subsection{The Klein--Gordon wave function formalism}

  As in the classical theory, conditioning ${\SPvec{A}}^\sharp(t,\SPvec{s},\SPvec{S})$
and ${A}^\sharp(t,\SPvec{s},\SPvec{S})$ with $\SPvec{s}=\SPvec{s}_k$ 
for each $k=1,...,N$ gives the $t$-synchronized $\widetilde{A}_k$ 
and $\widetilde{\SPvec{A}}_k$ (etc.) fields on $\RR\times\RR^{3N}_{\neq}$, 
\bea
&&\widetilde{A}_k
(t_1,{\SPvec{s}}_1,...,t_N,{\SPvec{s}}_N)\big|_{t_1=t_2=...=t_N=t}\big.
\equiv 
{A}_k(t,\SPvec{S})
\\
&&
\widetilde{\SPvec{A}}_k
(t_1,{\SPvec{s}}_1,...,t_N,{\SPvec{s}}_N)\big|_{t_1=t_2=...=t_N=t}\big. 
\equiv 
\SPvec{A}_k(t,\SPvec{S}).
\eea
  So far everything that has been stated in standard  space and time decomposition 
is exactly the same as in the classical theory. 
  The new material starts next, revealing that the Hamilton--Jacobi
equation has been replaced by a Klein--Gordon equation. 
   We first consider the single electron theory, then the generalization to many electrons.

	\subsubsection{A single electron}

 If there is only a single (positive or negative) electron in the world, 
no synchronization is necessary for $\widetilde\Phi$, i.e. 
$\widetilde\Phi(t,\SPvec{s}_1)\equiv \Phi(t,\SPvec{s}_1)$.
 Then, in standard space and time decomposition, 
\refeq{eq:GEOhamjacPDErecalib} becomes
\beq
- (- {\partial} \Phi -\plumi {\alpha}A_1)^2
+ {|{\nabla_1} \Phi -\plumi  {\alpha} {\SPvec{A}_1}|}^2 
+ 1 
- \varrho^{\,-1/2}{\mathbf{\Box}}_1 \varrho^{\,1/2}
= 0
\,,
\label{eq:SINGLEelectronHamJacFishEQ}
\eeq
with ${\mathbf{\Box}}_1 = -\partial^2 + \nabla^2_1$,
and \refeq{eq:kTHelectronContinuityEQrecalib} becomes
\beq
{\partial}
\Bigl(
\varrho \left(- {\partial} \Phi -\plumi {\alpha}A_1\right) 
\Bigr) 
+ {\nabla}_1\cdot 
\Bigl(
      \varrho\,( {\nabla_1} \Phi -\plumi {\alpha} {\SPvec{A}_1})^{}_{}
\Bigr) 
= 
0\,.
\label{eq:SINGLEelectronContinuityEQ}
\eeq
 Everywhere in the interior of $\supp(\varrho)$, we now multiply 
\refeq{eq:SINGLEelectronHamJacFishEQ} by $\varrho^{\,1/2} e^{i \Phi }$, 
and \refeq{eq:SINGLEelectronContinuityEQ} by $i\varrho^{\,-1/2} e^{i \Phi }$,
then add the so multiplied equations. 
  The result is a single complex \emph{linear} partial differential equation for 
\beq
\varrho^{\,1/2} e^{i \Phi }
\equiv 
\psi 
\,,
\label{eq:SINGLEelectronPSIofRHOphi}
\eeq
known as the Klein--Gordon equation
\beq
-(i{\partial} -\plumi {\alpha}A_1)^2\psi
+ {(-i{\nabla_1} -\plumi {\alpha} {\SPvec{A}_1})}^2\psi 
+ \psi
= 
0
\,,
\label{eq:KleinGordonSINGLEelectronEQ}
\eeq
which is coupled self-consistently to the total electromagnetic $^\sharp$potentials, 
and here restricted to the interior of $\supp(\varrho)$.
  Conversely, inserting \refeq{eq:SINGLEelectronPSIofRHOphi} into
\refeq{eq:KleinGordonSINGLEelectronEQ} and sorting into real and 
imaginary parts gives back the pair of equations
\refeq{eq:SINGLEelectronHamJacFishEQ}
and
\refeq{eq:SINGLEelectronContinuityEQ}. 
  Note that $|\psi|^2 = \varrho$ is a relativistic scalar, hence \emph{not} 
a probability density, as could have seemed by the resemblance with Born's
statistical law that $|\Psi|^2(t,\SPvec{s}_1)$ is a probability density for 
$\SPvec{s}_1$ when  $\Psi(t,\SPvec{s}_1)$ is the Schr\"odinger wave function.

 We next define the formal probability density field 
$\varrho \left(- {\partial} \Phi -\plumi {\alpha}A_1\right)\equiv \rho$ 
and the probability current vector-density field
$\varrho\,( {\nabla_1} \Phi -\plumi {\alpha} {\SPvec{A}_1})^{}_{}\equiv\SPvec{j}^{\mathrm{qu}}$
on configuration space, in terms of which \refeq{eq:SINGLEelectronContinuityEQ} takes the
familiar appearance of a continuity equation, 
\beq
{\partial} \rho + {\nabla_1}\cdot {\SPvec{j}^{\mathrm{qu}}}
= 
0.
\label{eq:SINGLEelectronContinuityEQjrho} 
\eeq
 But having identified the evolution equations for $\varrho$ and $\Phi$ with the Klein--Gordon
equation for $\psi$, we can express $\rho(t,\SPvec{s}_1)$ and ${\SPvec{j}^{\mathrm{qu}}}(t,\SPvec{s}_1)$ 
directly in terms of $\psi(t,\SPvec{s}_1)$, i.e.
\bea
\rho
\!\!\!&=&\!\!\!
\Im \left( \overline{\psi}\left(-{\partial} - \plumi i \alpha A_1\right)\psi\right)
\label{eq:QrhoOFpsi}
\,,
\\
{\SPvec{j}^{\mathrm{qu}}}
\!\!\!&=&\!\!\!
{\Im \left(\overline{\psi}\left( {\nabla}_1 - \plumi i \alpha \SPvec{A}_1\right)\psi\right)}
\label{eq:QcurrentOFpsi}
\,,
\eea
where $\Im$ means imaginary part, and we used the mathematical convention $\overline{\psi}$ for the 
complex conjugate of $\psi$ to avoid confusion with the star symbol for Hodge duals. 
 The r.h.s. of \refeq{eq:QrhoOFpsi} 
and the r.h.s. of \refeq{eq:QcurrentOFpsi} 
are recognized as the familiar expressions for the probability density and probability 
current density associated with the Klein--Gordon equation; notions that make sense
as long as $\rho \geq 0$.

 The ratio ${\SPvec{j}^{\mathrm{qu}}}(t,\SPvec{s}_1)/\rho(t,\SPvec{s}_1)$ 
defines the electron's quantum velocity field $\SPvec{v}_1^{\mathrm{qu}}(t,\SPvec{s}_1)$,
\beq
{\SPvec{v}_1^{\mathrm{qu}}}
\equiv
\frac{\Im \left(\overline{\psi}\left( {\nabla}_1 - \plumi i \alpha \SPvec{A}_1\right)\psi\right)}
     {\Im \left(\overline{\psi}\left(-{\partial} - \plumi i \alpha A_1\right)\psi\right)}
\label{eq:QvelocityOFpsi}
\,.
\eeq
 We remark that the  $\overline{\psi}$ in \refeq{eq:QvelocityOFpsi} 
can be replaced with $\psi^{-1}$ simultaneously in numerator and denominator
of the r.h.s. of \refeq{eq:QvelocityOFpsi}, which is readily checked.
 We next note that rewritten in terms of $\rho$ and ${\SPvec{v}_1^{\mathrm{qu}}}$, 
\refeq{eq:SINGLEelectronContinuityEQjrho} becomes 
\beq
{\partial} \rho + {\nabla_1}\cdot 
\left(\rho\, {\SPvec{v}_1^{\mathrm{qu}}}\right) 
= 
0\,,
\label{eq:QcontinuityEQrhov}
\eeq
and \refeq{eq:QcontinuityEQrhov} implies that the initial $\rho_0$ is being
transported by the velocity field ${\SPvec{v}_1^{\mathrm{qu}}}$ so that it 
would seem to follow that $\rho$ stays non-negative if it was so initially.
 In principle, however, `pathologies' of $ {\SPvec{v}_1^{\mathrm{qu}}}$
may develop, a priori speaking.

 In particular, the identification of 
$\SPvec{s}_1\mapsto\SPvec{v}_1^{\mathrm{qu}}(t,\SPvec{s}_1)$
with a \emph{velocity} field raises the question whether 
$\SPvec{v}_1^{\mathrm{qu}}(t,\,.\,)$ remains subluminal if it is
so initially, which is the case iff the Minkowski co-vector 
$\Im \left(\overline{\psi}\left( \dQ_1 - \plumi i \alpha \AQ_1\right)\psi\right)(t,\,.\,)$
remains time-like  at all $s_1$.
 Of course, if subluminality holds initially, one can argue  that by continuity
subluminality extends locally into the future, but whether this
holds for all future times is not a-priori clear. 
 We note that even if eventually $|\SPvec{v}_1^{\mathrm{qu}}(t,\,.\,)|>1$
somewhere, the obvious next question is whether the actual point motion generated 
by $\SPvec{v}_1^{\mathrm{qu}}(t,\,.\,)$ stays subluminal or whether it reaches 
those  regions where  $\SPvec{v}_1^{\mathrm{qu}}(t,\,.\,)$ is superluminal.

	\subsubsection{Many electrons}

 The many-electrons wave function formalism compounds the difficulties that
the second time derivatives of the wave operators brings with it.
 Here we only comment very briefly on the many-electrons wave function formalism,
leaving more detailed discussions for some future works. 

 To begin with, simply repeating the steps of the one-electron formalism now
for the $N$ particles situation, one concludes that in space and time decomposition 
the formal many-times functions $\widetilde\Phi(...,t_k,{\SPvec{s}_k,...})$ 
and $\widetilde\varrho(...,t_k,{\SPvec{s}_k,...})$
combine into the formal many-times wave function
\beq
\widetilde\psi
  \equiv 
{\widetilde\varrho\,}^{\frac{1}{2}} e^{i \widetilde\Phi}
\,,
\label{eq:kMANYelectronPSIofRHOphi}
\eeq
which has to satisfy $N$ many-times Klein--Gordon equations 
\beq
-(i{\partial}_k -\pmk {\alpha}\widetilde{A}_k)^2\widetilde\psi
+ {(-i{\nabla_k} -\pmk {\alpha} \widetilde{\SPvec{A}}_k)}^2\widetilde\psi
+ \widetilde\psi
= 
0
\,.
\label{eq:KleinGordonMANYelectronEQk}
\eeq
 As in the classical theory, without further restriction the $N$ equations 
\refeq{eq:KleinGordonMANYelectronEQk} would overdetermine $\widetilde\psi$, 
and so one should instead consider  a wave function
$\psi(t,\SPvec{S}) = \widetilde\psi(t,\SPvec{s}_1,...,t,\SPvec{s}_k,...)$
restricted to $t$-synchronized $\MM^{4N}_{\neq}$ 
($\cong \RR\times\RR^{3N}_{\neq}\subset\MM^{4N}_{\neq}$).
  However, since \refeq{eq:KleinGordonMANYelectronEQk} is of second order in
time, it is less obvious now than in the single-electron case what the
evolution equation for $\psi$ should be; thus,
$\partial^2\psi(t,\SPvec{S})$ involves mixed time derivatives of 
$\widetilde\psi(t,\SPvec{s}_1,...,t,\SPvec{s}_k,...)$, which are not 
determined by \refeq{eq:KleinGordonMANYelectronEQk}.
 In the classical setting this problem does not arise because the equations
for the classical $\widetilde\Phi$ are of first order in time. 
 Moreover, the additional problem arises that the obvious many-particles
analogue of the velocity formula \refeq{eq:QvelocityOFpsi} that comes to 
mind, namely $N$ ``quantum velocities'' fields
\beq
\SPvec{v}_k^{\mathrm{qu}}(t,\SPvec{S})
=
\frac{{\Im \left(\overline{\widetilde\psi}\left( {\nabla}_k 
                        - i \pmk  \alpha \widetilde{\SPvec{A}}_k\right)\widetilde\psi\right)}}
     {{\Im \left(\overline{\widetilde\psi}\left(-{\partial}_k 
                        - i \pmk  \alpha \widetilde{A}_k\right)\widetilde\psi\right)}}
\,,\label{eq:kQvelocityOFpsi}
\eeq
is not acceptable,\footnote{Note added 03/08/2004: 
                            Generally speaking, that is. In the decoherent approximation, variables
			    separate and one can work with \refeq{eq:KleinGordonMANYelectronEQk}
			    and \refeq{eq:kQvelocityOFpsi} as in the single-electron case.} 
for it does not lead to a continuity equation for any reasonable 
choice of probability density $\rho$ on configuration space $\RR^{3N}_{\neq}$.
 The following three options are possible ways out of the dilemma.

 First, in the context of test particle theory with given external fields, a many-electron
Klein--Gordon formalism has been worked out 
       \cite{rodi}
which operates with a single $\rho$ and $N$ currents ${\SPvec{j}_k^{\mathrm{qu}'}}$ defined in 
terms of higher derivatives of $\widetilde\psi$, and this formalism ought to be adaptable to 
our situation with total fields instead of externally given fields.
 Second, since the troubles come from having second-order time derivatives in the Klein--Gordon 
equation, one might think of using the familiar ``square-root Klein--Gordon'' equations instead.
 However, since already at the one-electron level such problems led Dirac to the invention of 
his first-order Dirac equation, one may want to take the above mentioned difficulties as an 
incentive to wait no longer but to now incorporate spin into the formalism, especially since 
this is the direction one will pursue eventually anyhow.
 Third, and last, one may want to take the formal relativistic many Klein--Gordon equations formalism
as a heuristic starting point for a non-relativistic approximation, which essentially consists 
of (i) replacing the square of the first-order co-variant time derivative in 
\refeq{eq:KleinGordonMANYelectronEQk} by twice the first-order co-variant time derivative itself, 
(ii) Born's law for $\rho$, i.e. $\rho\equiv |\widetilde\psi|^2$, through which the 
r.h.s. of the formal velocity field \refeq{eq:kQvelocityOFpsi} gets replaced by 
$\Im \left(\widetilde\psi^{-1}( {\nabla}_k- i \pmk  \alpha \widetilde{\SPvec{A}}_k)\widetilde\psi\right)$.
 Synchronization on $\RR\times\RR^{3N}_{\neq}$ now does not run into any problems, and
one obtains a many-body Schr\"odinger equation with potentials determined by the 
Maxwell--Born--Infeld $^\sharp$field equations, the sources of which move according
to the  de Broglie--Bohm velocity field. 
 Moreover, even though spin is not implemented then, the Pauli principle for 
fermions 
          \cite{streaterwightmanBOOK}
can of course be implemented now, but also vindicated, as discussed in 
           \cite{duerretalC}.
  While we will work out the non-relativistic approximation explicitly only in the one-electron
setting, the formal adaption of this to the $N$ electrons setting is then indeed straightforward.
 
	\subsection{The Cauchy problem}
 
 The Cauchy problem of the charge-free situation is identical to the 
one of the classical theory, and need not be repeated here.
 Also when point charges are present, part of the Cauchy problem is still 
the same, too, namely the Cauchy problem for the $^\sharp$fields; however, 
the part of the Cauchy problem dealing with the fields on configuration space  
has changed radically. 
 We only address the single electron version.

	\subsubsection{The configuration space problem}

 While in the classical single electron theory the Cauchy problem on 
configuration space dealt with a single field, $\Phi$, now we have
two fields, $\Phi$ and $\varrho$, or which is the same, one complex
field $\psi$. 
 This in itself is not a truly dramatic change, for also 
in the classical theory we could have added some luxury and 
amended $\Phi$ by a  \emph{passive} scalar amplitude field $\varrho$
satisfying \refeq{eq:kTHelectronContinuityEQrecalib} on the chosen
foliation, which on the standard foliation ($t$-synchronization)
becomes \refeq{eq:SINGLEelectronContinuityEQ}. 
 The radical change thus comes about not from having another field
$\varrho$ and \refeq{eq:SINGLEelectronContinuityEQ}, 
the radical change comes about through the \emph{second}-order
term $\varrho^{-1/2}{\mathbf{\Box}}_1 \varrho^{1/2}$ in 
\refeq{eq:SINGLEelectronHamJacFishEQ}.
 Indeed, in the amended-amplitude classical Hamilton--Jacobi theory
the fact that the Hamilton--Jacobi equation for $\Phi$ is of first order 
in time allows one to eliminate the time derivatives of $\Phi$ in 
\refeq{eq:SINGLEelectronContinuityEQ}, as a consequence of which  
\refeq{eq:SINGLEelectronContinuityEQ} in the classical amended
single electron theory is a first-order equation for $\varrho$, 
given that $\Phi$ satisfies the Hamilton--Jacobi PDE.
 In sharp contrast, in our single electron quantum theory the Cauchy problem 
is of second order for both $\Phi$ and $\varrho$; the highest time derivative 
of $\varrho$ occurs now in \refeq{eq:SINGLEelectronHamJacFishEQ}, the highest 
time derivative of $\Phi$ occurs now in \refeq{eq:SINGLEelectronContinuityEQ}.
 This turns the import of the two equations upside down.

 We note that equation \refeq{eq:QvelocityOFpsi} allows us to couple the configuration 
space indexed family of Maxwell--Born--Infeld $^\sharp$field equations
with point sources directly to the Klein--Gordon equation, 
without having any recourse whatsoever to $\varrho$ and $\Phi$.
 We remark that any solution $\psi$ which does not develop a zero in the interior 
of its support then maps into a unique global solution 
pair\footnote{We are not aware of general results as to 
      which initial conditions for the Klein--Gordon equation lead to 
      zeros of $\psi$ and which do not when the potentials $A_1$ and $\SPvec{A}_1$ 
      are given, not to speak of the self-consistent situation in which the
      potentials have to be solved for simultaneously with $\psi$.}
$\varrho,\Phi$
of \refeq{eq:SINGLEelectronContinuityEQ} and  \refeq{eq:SINGLEelectronHamJacFishEQ},
as verified by retracing backward the steps that led us to the Klein--Gordon equation.

 The Klein--Gordon equation \refeq{eq:KleinGordonSINGLEelectronEQ},
as a second order equation in time, requires initial data $\psi(0,\,.\,)$ 
and  $\partial\psi(0,\,.\,)$. 
 The fields $A_1$ and $\SPvec{A}_1$ that enter \refeq{eq:KleinGordonSINGLEelectronEQ}
are obtained by conditioning from the respective $^\sharp$fields with $\SPvec{s}=\SPvec{s}_1$;
the $^\sharp$fields in turn satisfy the first order equations
\refeq{eq:FOLIeqAscalSHARP}-\refeq{eq:FOLIpointSOURCESjjjSHARP}, with 
$\SPvec{v}_1^{\mathrm{qu}}(t,\SPvec{s}_1)$ given in 
\refeq{eq:QvelocityOFpsi}. 
 Initial data for all evolution equations have to be given. 
 It is straightforward to check that the initial value problem for the 
Maxwell--Born--Infeld $^\sharp$field equations with point sources 
moving according to the velocity field \refeq{eq:QvelocityOFpsi} 
and with the wave function $\psi$ satisfying the Klein--Gordon 
equation is well-defined.
 Whether the Cauchy problem leads to global or just local existence and uniqueness results 
is an interesting open problem.

 The Cauchy problem described is autonomous in the sense that the actual
electromagnetic spacetime does not figure. 
 Of course, to obtain the actual electromagnetic spacetime from a solution of 
the $^\sharp$fields - Klein--Gordon equations, data for the
$^\sharp$fields have to reduce to the data for the actual fields 
when the actual particle configuration is substituted for the generic one,
and data for $\psi$ and $\partial\psi$ need to give the actual 
initial velocity of the point charge.

	\subsubsection{The actual motion and the actual fields}

 Once $\SPvec{v}_1^{\mathrm{qu}}(t,\SPvec{s}_1)$ has been computed autonomously
by solving the coupled system of Klein--Gordon and $^\sharp$field equations, 
one finally can solve the relativistic de Broglie--Bohm type guiding equation 
with given initial data $\SPvec{r}_1(0)$ to obtain the actual trajectory 
$t\mapsto \SPvec{r}_1(t)$ of the point charge, i.e. $\SPvec{r}_1(t)$ satisfies
\beq
\bulldif{\SPvec{r}}_1(t)= \SPvec{v}_1^{\mathrm{qu}}(t,\SPvec{r}_1(t))
\,.
\label{eq:dBBeq}
\eeq
 Once this has been done, the actual point sources in $\MM^4$ are known, too, 
and given by $j(t,\SPvec{s})= \plumi \delta_{\SPvec{r}_1(t)}(\SPvec{s})$
and
$\SPvec{j}(t,\SPvec{s})= \plumi \delta_{\SPvec{r}_1(t)}(\SPvec{s})\bulldif{\SPvec{r}}_1(t)$,
with ${\SPvec{r}}_1(t)$ satisfying \refeq{eq:dBBeq}. 
 These are indeed the familiar expressions 
\refeq{eq:FOLIpointSOURCESrho} and \refeq{eq:FOLIpointSOURCESjjj} for the 
electric `density' and electric current `vector density' of 
a single point charge at $\SPvec{s}_1 = \SPvec{r}_1(t)$ moving with velocity 
$\bulldif{\SPvec{r}}_1(t)$.
 It is straightforward to verify that charge conservation is guaranteed.

  This fully vindicates our designation of $\SPvec{v}_1^{\mathrm{qu}}(t,\SPvec{r}_1(t))$ 
as the  velocity of the electron in the quantum theory.

 Having the actual point-charge source terms for the actual Maxwell--Born--Infeld
field equations, we could now solve them bottom-up to get the actual fields; 
however, as already emphasized, by the very setup of the theory we can simply
substitute the actual position $\SPvec{r}_1(t)$ for the generic $\SPvec{s}_1$ 
in the $^\sharp$fields to obtain the actual Maxwell--Born--Infeld fields.
 Thus, $\SPvec{A}(t,\SPvec{s}) = \,{^\sharp}\!\SPvec{A}(t,\SPvec{s},\SPvec{r}_1(t))$,
etc., which solves the actual Maxwell--Born--Infeld field equations top-down.

      \section{Application to atoms}

  While the absence of spin in the Klein--Gordon equation and the absence
of photons from the electromagnetic fields limit the applicability of our
theory in practical situations, we do get the correct low-energy physics 
whenever spin effects and the photonic nature of the electromagnetic fields 
are known to contribute only small corrections.  
 Thus, in the non-relativistic limit we obtain the correct Schr\"odinger equation 
with Coulomb interaction.
 Since in our theory the electromagnetic fields are the total fields, and the 
self-field energies are all finite, the Coulomb interactions emerge in the 
non-relativistic limit \emph{without} any truncation and renormalization. 
 We shall  work this out explicitly for the hydrogen atom, a
non-genuinely electromagnetic example for which it can be assumed that
the nucleus and the electron move at non-relativistic speeds.
 To keep the presentation as simple as possible, we actually treat the nucleus 
in the Born--Oppenheimer approximation as infinitely massive; we also assume the 
nucleus to be a point without magnetic moment.
 The extensions of all the genuinely electromagnetic formulas to this 
non-genuine setting are straightforward.
 The many-electrons atom with nuclear charge $z>1$ and $N=z$ negative electrons
will be treated elsewhere.

 To leading order in an expansion in terms of powers of $\alpha$,
assumed to be small, the familiar data of non-relativistic quantum theory emerge 
in the formal limit $\beta\downarrow 0$ if and only if we identify $\alpha$ with 
Sommerfeld's fine structure constant --- as we have argued non-rigorously already in 
   \cite{KiePapI}.
 A non-vanishing Born's aether constant $\beta$ in turn induces corrections to the 
spectrum, which must be small, and this puts some rough and ready
upper bounds on $\beta$. 
  
  \subsection{The hydrogen atom}

  For the hydrogen atom, in Born--Oppenheimer approximation, the infinitely 
massive point nucleus of charge $+1$ can be assumed to be at rest at the origin $\SPvec{0}$ of
our space.
 This requires adding a term $\delta_{\SPvec{o}}(\SPvec{s})$ to the charge density.
 The single point electron moves along the trajectory $t\mapsto\SPvec{s}_1 = \SPvec{r}_1(t)$
(henceforth, we drop the suffix $_1$ from $\SPvec{r}_1(t)$),
and the charge density and current vector density then read 
 \bea
        j (t,{\SPvec{s}})
\!\!\!&=&\!\!\!
 \delta_{{\SPvec{o}}}({\SPvec{s}})
-
 \delta_{{\SPvec{r}}(t)}({\SPvec{s}})
\label{eq:hydroCHARGE}
\\
        {\SPvec{j}}(t,{\SPvec{s}})
\!\!\!&=&\!\!\!
-
 \delta_{{\SPvec{r}}(t)}({\SPvec{s}})\bulldif{\SPvec{r}}(t)
\, .
\quad\qquad
\label{eq:hydroCURRENT}
\eea
 These are now the point source terms for the Maxwell--Born--Infeld field 
equations, which are supplemented by the asymptotic conditions that
all fields vanish at spatial infinity. 
 The total electric and magnetic fields have potentials which in turn enter the 
Klein--Gordon equation.
 More precisely, what enters the Klein--Gordon equation are not the actual fields for
the unknown actual position and velocity, but the conditioned $^\sharp$fields 
for the generic positions on configuration space and the velocities associated 
to them by the velocity field $\SPvec{v}^{\mathrm{qu}}$.
  Solutions of the Klein--Gordon equation
\refeq{eq:KleinGordonSINGLEelectronEQ} on single-electron configuration
space define the velocity vector field 
\refeq{eq:QvelocityOFpsi} on that configuration space, which evolves 
any actual electron's position vector $\SPvec{r}(t)$ via the relativistic 
de Broglie--Bohm type guiding equation 
$\bulldif{\SPvec{r}}(t)= \SPvec{v}^{\mathrm{qu}}(t,\SPvec{r}(t))$,
which in turn determines, for each actual trajectory the electron traces out, 
the point source terms \refeq{eq:hydroCHARGE} and \refeq{eq:hydroCURRENT}
for the Maxwell--Born--Infeld field equations, closing the loop. 
 However, since the  Maxwell--Born--Infeld $^\sharp$field equations 
have to be solved along with the Klein--Gordon equation, the actual 
Maxwell--Born--Infeld field equations need not to be solved again;
their solution is simply obtained then by substituting the actual 
trajectory for the generic one in the respective solutions of the
$^\sharp$fields equations.

\subsubsection{Existence of infinitely many radiation-free bound states.}

 We first establish the existence of stationary solutions of our coupled 
system of equations.
 Stationarity in the Lorentz gauge with asymptotically (at spatial
infinity) vanishing conditions for the $^\sharp$potentials means that the
electric potential 
$^\sharp{A}(t,\SPvec{s},\SPvec{s}_1) \equiv \,{^\sharp}\!A_0(\SPvec{s},\SPvec{s}_1)$,
and the magnetic vector potential 
$^\sharp\SPvec{A}(t,\SPvec{s},\SPvec{s}_1) \equiv \,{^\sharp}\!\SPvec{A}_0(\SPvec{s},\SPvec{s}_1)$.
 The potential terms in the Klein--Gordon equation are then explicitly time-independent, 
and the only time-dependence allowed is in an overall phase rotation of the wave function, 
thus $\psi(t,\SPvec{s}_1) = e^{-i {\veps}t}\psi^{\mathrm{stat}}(\SPvec{s}_1)$. 
 This in turn implies $\SPvec{v}^{\mathrm{qu}}(t,\SPvec{s}_1)\equiv \SPvec{0}$,
which implies $^\sharp\SPvec{j}(t) =\SPvec{0}$ (wherever $\SPvec{s}_1$ may be),
and this now implies 
$^\sharp{j}(t,{\SPvec{s},\SPvec{s}_1})
= \delta_{{\SPvec{o}}}({\SPvec{s}})- \delta_{{\SPvec{s}}_1}({\SPvec{s}})$.
 Having static sources together with $\partial\,{^\sharp}\!\SPvec{D}\equiv \SPvec{0}$ 
implies $\nabla\times\,{^\sharp}\SPvec{H}\equiv \SPvec{0}$, whence 
$^\sharp\SPvec{H}\equiv \SPvec{0}$,
whence  $^\sharp\SPvec{B}\equiv \SPvec{0}$, 
and therefore $^\sharp\SPvec{A}\equiv \SPvec{0}$. 

 Hence, the only allowed fields are electrostatic, and
we need to solve the electrostatic Coulomb--Born--Infeld equation
\beq
-\nabla \cdot
\frac{ \nabla{\,{^\sharp}\!A_0}(\SPvec{s},\SPvec{s}_1)}
{\sqrt{1 - \beta^4| \nabla{\,{^\sharp}\!A_0}(\SPvec{s},\SPvec{s}_1) |^2 }}
= 
   4 \pi \delta_{{\SPvec{o}}}({\SPvec{s}})-    4 \pi\delta_{{\SPvec{s}}_1}({\SPvec{s}})
\label{eq:FOLIeqAstaticHYDRO}
\, 
\eeq 
for arbitrary location $\SPvec{s}_1$ of the electron, 
with asymptotic condition ${\,{^\sharp}\!A}_0({\SPvec{s}},\SPvec{s}_1)\to 0$ as 
$|{\SPvec{s}}|\to\infty$.
 Any such solution of \refeq{eq:FOLIeqAstaticHYDRO} is unique, see our proof in
   \cite{KiePapI}.
 As for the existence of solutions, we invoke Bartnik's remark in 
   \cite{bartnik}
that his existence Theorem 5.4 generalizes to maximal space-like slices with light cone
singularities to anticipate the existence of electrostatic potentials with two
point charges for any configuration of the electron and nucleus positions;
however, an explicit existence proof should be supplied eventually.

 Of the solution 
$\SPvec{s}\mapsto \,{^\sharp}\!A_0(\SPvec{s},\SPvec{s}_1)$ 
to \refeq{eq:FOLIeqAstaticHYDRO}
only $ \,{^\sharp}A_0(\SPvec{s}_1,\SPvec{s}_1)\equiv A_0(\SPvec{s}_1)$ 
is needed in the Klein--Gordon equation.\footnote{Strictly speaking, to keep with our notational conventions, 
                                                  instead of $A_0(\SPvec{s}_1)$ we should write $A_{1,0}(\SPvec{s}_1)$,
						  but no confusion should arise  from dropping the suffix 1 here.}
 Interestingly enough, $A_0(\SPvec{s}_1)$ can be calculated without 
knowledge of the complete solution 
$\SPvec{s}\mapsto \,{^\sharp}\!A_0(\SPvec{s},\SPvec{s}_1)$ of \refeq{eq:FOLIeqAstaticHYDRO},
as we show next.

 For the purpose of calculating $A_0(\SPvec{s}_1)$, we
remark that the solution to \refeq{eq:FOLIeqAstaticHYDRO} must satisfy
\beq
-\frac{ \nabla{\,{^\sharp}\!A}_0 }{ \sqrt{ 1 - \beta^4| \nabla{\,{^\sharp}\!A_0}|^2 }}
= 
\,{^\sharp}\!{\SPvec{D}}^{(2)}_{\mathrm{Coulomb}} 
+
\nabla\times \,{^\sharp}\!\SPvec{Z}
\label{eq:FOLIeqAofZstatic}
\, 
\eeq
for all $\SPvec{s}\neq \SPvec{0}$ or $\SPvec{s}_1$, where 
\beq
\,{^\sharp}\!{\SPvec{D}}^{(2)}_{\mathrm{Coulomb}} ({\SPvec{s}},\SPvec{s}_1)
=
-\nabla \Big(|\SPvec{s}|^{-1} -{|\SPvec{s}-\SPvec{s}_1|^{-1}}\Big)
\label{eq:FOLIeqDzweiCOULOMB}
\,,
\eeq
so that for a given solution $\,{^\sharp}\!A_0$ of \refeq{eq:FOLIeqAstaticHYDRO}, 
$\nabla \times \,{^\sharp}\!{\SPvec{Z}}$ 
is uniquely defined by \refeq{eq:FOLIeqAofZstatic}
for all $\SPvec{s}\neq \SPvec{0}$ or $\SPvec{s}_1$; 
and since $\beta^2| \nabla{\,{^\sharp}\!A_0}({\SPvec{s}},{\SPvec{s}_1})|\to 1$
when $\SPvec{s}\to \SPvec{0}$ or $\SPvec{s}_1$, we can extend $\nabla\times \,{^\sharp}\!\SPvec{Z}$ continuously to 
all $\SPvec{s}\in\RR^3$ by setting $\nabla\times \,{^\sharp}\!\SPvec{Z}(\SPvec{s},\SPvec{s}_1) =\SPvec{0}$ 
for $\SPvec{s}= \SPvec{0},\ \SPvec{s}_1$.
 The field ${\SPvec{s}}\mapsto\,{^\sharp}\!\SPvec{Z}({\SPvec{s}},\SPvec{s}_1)$ 
(with $\SPvec{s}_1$ as parameter) is itself an electrostatic vector potential 
which we can assume to vanish for $|{\SPvec{s}}|\to\infty$. 
 Note that $\,{^\sharp}\!\SPvec{Z}$ is defined only up to the gauge transformation 
$\,{^\sharp}\!{\SPvec{Z}}\to \,{^\sharp}\!{\SPvec{Z}} + 
\nabla \,{^\sharp}U$, under which \refeq{eq:FOLIeqAofZstatic} is invariant. 
  We can remove this freedom by imposing the gauge condition 
$
        {\nabla}\cdot  \,{^\sharp}\!{\SPvec{Z}}
=
        0
$,
which can always be achieved by solving a Poisson equation for $\,{^\sharp}U$, if necessary.
 However, what matters is only $\nabla\times\,{^\sharp}\!\SPvec{Z}$.
 Easily inverting \refeq{eq:FOLIeqAofZstatic} we get
\beq
-\nabla \,{^\sharp}\!A_0
= 
\frac{
     \,{^\sharp}\!{\SPvec{D}}^{(2)}_{\mathrm{Coulomb}} + \nabla \times \,{^\sharp}\!\SPvec{Z}
     }{
      \sqrt{    1 + \beta^4| \,{^\sharp}\!{\SPvec{D}}^{(2)}_{\mathrm{Coulomb}} + \nabla \times \,{^\sharp}\!\SPvec{Z}|^2
     }}
\label{eq:FOLIeqAofZstaticREVERSE}
\,,
\eeq 
and integration along any path from $\SPvec{s}_0$ to $\SPvec{s}$ gives the identity
${^\sharp}\!A_0(\SPvec{s},\SPvec{s}_1) = {^\sharp}\!A_0(\SPvec{s}_0,\SPvec{s}_1) 
+ \int_{\SPvec{s}_0}^{\SPvec{s}}\nabla \,{^\sharp}\!A_0(\hat\SPvec{s},\SPvec{s}_1)\cdot \dd\hat\SPvec{s}$, 
with the integrand given by the r.h.s. of \refeq{eq:FOLIeqAofZstaticREVERSE}.
 Of course, $\SPvec{s}_0$ should be picked conveniently so that 
$\,{^\sharp}\!A_0(\SPvec{s}_0,\SPvec{s}_1) = \SPvec{0}$;
for instance, the standard convention (valid for more than two point charges as well) would
be to let $\SPvec{s}_0\to \partial \RR^3$ (infinity).
 However, notice that by the symmetry of the problem we know that 
$\,{^\sharp}\!A_0(\SPvec{s}_0,\SPvec{s}_1) = \SPvec{0}$
for all 
$\SPvec{s}_0\in \{0.5 \SPvec{s}_1 + \SPvec{s}^\perp, \SPvec{s}^\perp\cdot\SPvec{s}_1 = \SPvec{0} \}$, 
and so we may want to pick such an $\SPvec{s}_0$. 
 Furthermore, a straightforward calculation shows that on the straight line joining the 
nuclear point charge (representing the proton) and the point electron, we have 
\beq
\nabla \times
\left.
\frac{
       \,{^\sharp}\!{\SPvec{D}}^{(2)}_{\mathrm{Coulomb}} (\SPvec{s},\SPvec{s}_1)
     }{
       \sqrt{    1 + \beta^4| \,{^\sharp}\!{\SPvec{D}}^{(2)}_{\mathrm{Coulomb}} (\SPvec{s},\SPvec{s}_1)|^2
     }}
\right|_{\SPvec{s}\in \{\xi \SPvec{s}_1;\, \xi\in\RR \}}
=
\SPvec{0}
\label{eq:FOLIeqAofZstaticREVERSEcurl}
\, ,
\eeq
so that we may in fact conclude that 
\beq
\nabla \times \,{^\sharp}\!\SPvec{Z}(\SPvec{s},\SPvec{s}_1)\Big|_{\SPvec{s}\in \{\xi \SPvec{s}_1;\, \xi\in\RR \}} 
=\SPvec{0}
\,.
\eeq
 Hence, picking $\SPvec{s}_0 = 0.5\SPvec{s}_1$, we find the explicit one-dimensional integral formula
\beq
A_0(\SPvec{s}_1) 
=  
-\int_{1/2}^{1}
\frac{
      \SPvec{s}_1 \cdot \,{^\sharp}\!{\SPvec{D}}^{(2)}_{\mathrm{Coulomb}} (\xi \SPvec{s}_1,\SPvec{s}_1)
     }{
       \sqrt{    1 + \beta^4| \,{^\sharp}\!{\SPvec{D}}^{(2)}_{\mathrm{Coulomb}} (\xi \SPvec{s}_1,\SPvec{s}_1)|^2
     }}
\,\dd\xi
\,,
\label{eq:AatSoneINTEGRALformula}
\eeq
\newpage

\noindent
as announced. 
 (By the symmetry of the configuration, the integral can be replaced by $1/2$ the same integral taken
in the limits from $0$ to $1$.)

 The integral \refeq{eq:AatSoneINTEGRALformula} can be manipulated into a form that shows a strong 
resemblance to integrals listed in 
     \cite{gradshteyn}
which can be evaluated in closed form with the help of elliptic integrals and elementary functions;
however, we have not yet succeeded to evaluate it in closed form.
 Therefore we have to resort to computing $A_0(\SPvec{s}_1)$ for small and large values of 
$|\SPvec{s}_1|/\beta$.
 To state the proposition, we recall that
\beq
A_{\mathrm{Born}}^\plumi({\SPvec{s}}) 
=  
\plumi \frac{1}{\beta}
\int_{|{\small\SPvec{s}}|/\beta}^\infty \frac{\dd{x}}{\sqrt{1+ x^4}} 
\,
\label{eq:BornsElectricPot}
\eeq
is Born's solution for the electrostatic potential of a single positive or negative electron 
at the origin of space.

\begin{Prop} 
\label{AnullASYMPTOTICS} 
 If the electron is near the nucleus, more precisely iff $|\SPvec{s}_1| < 2\sqrt{2}  \beta$, 
then $A_0(\SPvec{s}_1)$ can be expanded into a convergent series in powers of
$|\SPvec{s}_1|/\beta$, i.e.
\beq
A_0(\SPvec{s}_1) 
= 
- \frac{1}{ 2 \beta}\left[
                            \frac{|\SPvec{s}_1|}{\beta} 
			    -\frac{|\SPvec{s}_1|^5}{\beta^5} 
			    \int_{1/2}^{1}
	     \frac{\xi^4 (1-\xi)^4}{\left(1- 2\xi(1-\xi)\right)^2}
	                    \dd\xi
			    + O\left(\frac{|\SPvec{s}_1|^9}{\beta^9}\right) 
                       \right]
\,.
\label{eq:AatSoneINTEGRALformulaSMALLsONE}
\eeq
 If, on the other hand, the electron is far from the nucleus, i.e. for
$|\SPvec{s}_1| \geq 2\sqrt{2}  \beta$, then $A_0(\SPvec{s}_1)$ can be 
expanded (asymptotically exact for $|\SPvec{s}_1| \to \infty$) to get
\beq
A_0(\SPvec{s}_1) 
=
 A_{\mathrm{Born}}^{(-)}({\SPvec{0}}) 
+
\frac{1}{|\SPvec{s}_1|}\left[1 - U\left(\frac{\beta}{|\SPvec{s}_1|}\right)\right]
\,,
\label{eq:AatSoneASYMPTOTICS}
\eeq
with  $U(\beta/|\SPvec{s}_1|) < 0$ and
$|U(\beta/|\SPvec{s}_1|)| < 2 {\beta}/{|\SPvec{s}_1|}$ 
 for $|\SPvec{s}_1|$ large enough. 
\end{Prop}

\begin{Rema}$\!\!\!\!$ We can take \refeq{eq:AatSoneASYMPTOTICS} as defining 
$U(\beta/|\SPvec{s}_1|)$ for all $|\SPvec{s}_1|$, and indeed 
$U(\beta/|\SPvec{s}_1|)$ is then well defined for all $|\SPvec{s}_1|>0$. 
Also, using \refeq{eq:AatSoneINTEGRALformulaSMALLsONE}
one easily sees that then also 
$U(\beta/|\SPvec{s}_1|) = 
1 + |\SPvec{s}_1|  A_{\mathrm{Born}}^{(-)}({\SPvec{0}}) + \frac{1}{2}\frac{|\SPvec{s}_1|^2}{\beta^2} + ...$ 
for small $|\SPvec{s}_1|$.
 It follows that $U$ is bounded.
 It also follows that $U(\beta/|\SPvec{s}_1|) >0$ for small $|\SPvec{s}_1|$,  and since 
$U(\beta/|\SPvec{s}_1|) < 0$ for large $|\SPvec{s}_1|$, we believe that 
$U(\beta/|\SPvec{s}_1|) =0 $ for exactly one $|\SPvec{s}_1|$; 
however, so far we have not been able to prove this.
\end{Rema}

\newpage

\noindent
\textit{Proof of Proposition \ref{AnullASYMPTOTICS}}. 
 The integral formula \refeq{eq:AatSoneINTEGRALformula} can be easily
rendered as
\beq
A_0(\SPvec{s}_1) 
=
-
\frac{|\SPvec{s}_1|}{ \beta^2}
\int_{1/2}^{1}
           \left( 1+ \frac{|\SPvec{s}_1|^4}{ \beta^4}
	     \frac{\xi^4 (1-\xi)^4}{\left(1- 2\xi(1-\xi)\right)^2}\right)^{-1/2}
{\dd\xi}
\,.
\label{eq:AatSoneINTEGRALformulaEXPLICIT}
\eeq
 The Taylor expansion of \refeq{eq:AatSoneINTEGRALformulaEXPLICIT}
for $|\SPvec{s}_1| < 2\sqrt{2}  \beta$ is elementary and gives 
\refeq{eq:AatSoneINTEGRALformulaSMALLsONE}.
 Note that only powers $(|\SPvec{s}_1|/\beta)^{4k+1}$, $k=0,1,2,...$ enter,
and that all coefficients are integrals of bounded 
rational functions which can be evaluated in closed form.
 
 To obtain the asymptotic expansion of $A_0(\SPvec{s}_1)$ 
for $|\SPvec{s}_1|\to\infty$, we rewrite \refeq{eq:AatSoneINTEGRALformula} as
\beq
A_0(\SPvec{s}_1) 
=
\frac{1}{ \beta}
\int_{2\sqrt{2}\beta/|\SPvec{s}_1|}^{\infty}
\frac{f^\prime(y)}{\sqrt{1+ x^4}}
      \dd{x}
\,,
\label{eq:AnullATsONElargeINT}
\eeq
where $xy = \beta/|\SPvec{s}_1|$, and $f^\prime$ means the derivative of $f$, with 
\beq
f(y) = \sqrt{\viertel + y^2 - y\sqrt{1 + y^2}}
\,.
\label{eq:fDEFINITION}
\eeq
 Writing out $f^\prime(y)$ explicitly is not very illuminating.
 Fortunately, all we need are the\hfill
following features of the map $y\mapsto f^\prime(y)$, which are straightforward to prove: 

\textit{(i) $y\mapsto f^\prime(y) + 1$ is strictly negative on $(0,1/2\sqrt{2})$;}

\textit{(ii) $y\mapsto f^\prime(y)$ is decreasing on $[0,1/2\sqrt{2})$ and
strictly decreasing on $(0,1/2\sqrt{2})$;}

\textit{(iii) $y\mapsto f^\prime(y)$ is strictly concave on $[0,1/2\sqrt{2})$;}

\textit{(iv) $y\mapsto f^\prime(y)$ is analytic on $[0,1/2\sqrt{2})$, and 
the Taylor--MacLaurin expansion}

\textit{{\quad\ } about $y =0$ of $f^\prime(y)$ reads (with $y\geq{0}$)}
\beq
f^\prime(y) = -1 -\frac{3}{2}y^2 - 4 y^3 - \frac{75}{8}y^4 +O(y^5);
\eeq

\textit{(v) $y\mapsto f^\prime(y)$ is bounded above and below on $[0,1/2\sqrt{2})$ by}
\beq
-\frac{1}
      {\sqrt{3}} 
 \frac{1}
      {\sqrt{1-2\sqrt{2}y}}
\geq 
    f^\prime(y) 
\geq 
-\frac{1}
      {\sqrt{3}} 
 \frac{1+\frac{3}{2}(1 - 2\sqrt{2}y)}
      {\sqrt{1- 2\sqrt{2}y}} 
\,,
\eeq

\textit{{\quad\ } and the difference of left- and right-hand sides $\downarrow 0$ as $2\sqrt{2}y\uparrow 1$.}

\newpage
\noindent
 We now prove first the leading order asymptotics in 
\refeq{eq:AatSoneASYMPTOTICS}, i.e. we show that 
\beq
\lim_{{|\SPvec{s}_1|}\to\infty} 
A_0({\SPvec{s}}_1)
=
 A_{\mathrm{Born}}^{(-)}({\SPvec{0}})
\,.
\label{eq:AatSoneASYMPTOTICSlead}
\eeq
 Indeed, by (i) we have ${f^\prime(y)}\leq -1$. 
 Inserting this estimate in \refeq{eq:AnullATsONElargeINT} gives
\beq
A_0(\SPvec{s}_1) 
\leq 
- \frac{1}{ \beta}
  \int_{2\sqrt{2}\beta/|\SPvec{s}_1|}^{\infty}
    \frac{1}{\sqrt{1+ x^4}}
  \dd{x}
\,,
\label{eq:upperESTofAatSone}
\eeq
and taking the limit gives
\beq
\limsup_{{|\SPvec{s}_1|}\to\infty} 
A_0(\SPvec{s}_1) 
\leq 
 A_{\mathrm{Born}}^{(-)}({\SPvec{0}}) 
\,.
\label{eq:limsupESTofAatSone}
\eeq
  On the other hand, for any small $\eps >0$, if $x\geq\eps$ then $y\leq  \beta/(|\SPvec{s}_1|\eps)$;
hence, for fixed $\eps >0$ and any $x\geq\eps$ we have $\lim_{{|\SPvec{s}_1|}/\beta\to\infty} f^\prime(y)=-1$,
uniformly, and this gives us
\beq
\lim_{{|\SPvec{s}_1|}\to\infty} 
\frac{1}{ \beta}
\int_{\eps+2\sqrt{2}\beta/|\SPvec{s}_1|}^{\infty}
\frac{f^\prime(y)}{\sqrt{1+ x^4}}
      \dd{x}
=
- \frac{1}{ \beta}
     \int_{\eps}^{\infty}
\frac{1}{\sqrt{1+ x^4}}
      \dd{x}
\geq
 A_{\mathrm{Born}}^{(-)}({\SPvec{0}}) 
\,,
\label{eq:liminfESTofAatSoneA}
\eeq
while 
\bea
&&
\liminf_{{|\SPvec{s}_1|}\to\infty} 
\frac{1}{ \beta}
\int_{2\sqrt{2}\beta/|\SPvec{s}_1|}^{\eps+2\sqrt{2}\beta/|\SPvec{s}_1|}
\frac{f^\prime(y)}{\sqrt{1+ x^4}}
      \dd{x}
\nonumber
\\
&&\geq
\liminf_{{|\SPvec{s}_1|}\to\infty} 
-\frac{5}
      {2\sqrt{3}} 
 \frac{1}
      {\beta}
\int_{2\sqrt{2}\beta/|\SPvec{s}_1|}^{\eps+2\sqrt{2}\beta/|\SPvec{s}_1|}
 \frac{\sqrt{x}}
      {\sqrt{x- 2\sqrt{2}\beta/{|\SPvec{s}_1|}}} 
      \dd{x}
\nonumber
\\
&&\geq
\liminf_{{|\SPvec{s}_1|}\to\infty} 
-\frac{5}
      {\sqrt{3}} 
 \frac{1}
      {\beta}
{\sqrt{\eps}}
 \sqrt{\eps+2\sqrt{2}\frac{\beta}{|\SPvec{s}_1|}}
=
- \frac{5}
      {\sqrt{3}} 
 \frac {1}
      {\beta}{\eps}
\,
\label{eq:liminfESTofAatSoneB}
\eea
 for all $\eps$. 
 In \refeq{eq:liminfESTofAatSoneB}, we used the estimate $\sqrt{1+x^4} \geq 1$ 
in conjunction with (i), followed by the lower estimate for $f^\prime(y)$ in (v) 
(which we further estimated by discarding the positive $y$ contribution in the numerator), 
followed by an elementary integration by parts and the omission of a manifestly positive 
additive integral.
 Estimates \refeq{eq:limsupESTofAatSone} on the one hand, and 
\refeq{eq:liminfESTofAatSoneA} and \refeq{eq:liminfESTofAatSoneB}
on the other, establish \refeq{eq:AatSoneASYMPTOTICSlead}.

 To establish the next-to-leading order term, we now show that
\beq
\lim_{{|\SPvec{s}_1|}\to\infty} 
|{\SPvec{s}}_1|
\left(
A_0({\SPvec{s}}_1)
-
A_{\mathrm{Born}}^{(-)}({\SPvec{0}}) 
\right)
=
1.
\label{eq:AnullOFsONEasymptoticsNEXTtoLEAD}
\eeq
 We first notice that 
$
|{\SPvec{s}}_1|
\left(A_0(
{\SPvec{s}}_1)-A_{\mathrm{Born}}^{(-)}({\SPvec{0}})
\right)
$
depends on $\beta$ and $|{\SPvec{s}}_1|$ exclusively 
through the combination $\beta/|{\SPvec{s}}_1|$, and
since $\beta/|{\SPvec{s}}_1|\downarrow{0}$ as
$|{\SPvec{s}}_1|\uparrow\infty$, it is convenient here
to introduce the abbreviation $\beta/|{\SPvec{s}}_1|\equiv\zeta$.
The limit $|{\SPvec{s}}_1|\uparrow\infty$ then becomes the
limit $\zeta\downarrow{0}$.
 To carry out this limit we split the integral for 
$A_{\mathrm{Born}}^{(-)}({\SPvec{0}})$ in two,
\beq
A_{\mathrm{Born}}^{(-)}({\SPvec{0}}) 
=
-
 \frac{1}{ \beta}
  \int_0^{2\sqrt{2}\beta/|\SPvec{s}_1|}
    \frac{1}{\sqrt{1+ x^4}}
  \dd{x}
-
 \frac{1}{ \beta}
  \int_{2\sqrt{2}\beta/|\SPvec{s}_1|}^{\infty}
    \frac{1}{\sqrt{1+ x^4}}
  \dd{x}
\,,
\label{eq:AbornNULLsplit}
\eeq
then lump the (negative of the) second integral together with the
integral \refeq{eq:AnullATsONElargeINT} for $A_0({\SPvec{s}}_1)$,
while the (negative of the) first integral will be handled on its own.
 After multiplication of these integrals by $|{\SPvec{s}}_1|$, we change 
integration variables $x\to x/\zeta\equiv\xi$ in the (negative of the) first integral of 
\refeq{eq:AbornNULLsplit} and $x\to y\ (=1/\xi)$ in the `lumped' integral.
 After this reshuffling of terms, the l.h.s. in 
\refeq{eq:AnullOFsONEasymptoticsNEXTtoLEAD}
becomes a sum of two limits $\zeta\to{0}$ which can be carried out
easily using monotone convergence, 
\bea
&&\lim_{\zeta\to{0}}
\left(
  \int_0^{2\sqrt{2}}
\!\!
\frac{\dd\xi}{\sqrt{1+ \zeta^4\xi^4}}
+
  \int_0^{1/2\sqrt{2}}
\frac{1+f^\prime(y)}{\sqrt{1+ \zeta^4/ y^4}}
\frac{\dd{y}}{y^2} 
\right)
\nonumber
\\
&&\qquad
=
  \int_0^{2\sqrt{2}}
  \dd{\xi}
+
  \int_0^{1/2\sqrt{2}}
\frac{1+f^\prime(y)}{y^2} 
\dd{y}
\label{eq:AnullOFsONEasympNEXTtoLEADfinalINT}
\eea
 The first of these limiting integrals obviously equals ${2\sqrt{2}}$.
 As for the second limiting integral, note that by the Taylor series
of $f^\prime(y)$ about $y=0$ its integrand is regular at $y=0$, while 
the bounds in (v) show that the singularity at $y\uparrow 1/2\sqrt{2}$ 
is a reciprocal square root, hence the integrand is integrable there, too. 
 By (i) this integral is negative. 
 To evaluate this integral, we write it as a limit of an integral as the lower
limit of integration of that integral tends to null, which gives (after carrying
out one obvious integration)
\beq
  \int_0^{1/2\sqrt{2}}
\frac{1+f^\prime(y)}{y^2} 
  \dd{y}
=
-2\sqrt{2} + 
\lim_{\eps\downarrow{0}} 
\left(
\frac{1}{\eps} 
+
  \int_\eps^{1/2\sqrt{2}}
\frac{f^\prime(y)}{y^2} 
   \dd{y}
\right)
\,.
\label{eq:EPSdownlimitINT}
\eeq
 To evaluate the remaining integral in \refeq{eq:EPSdownlimitINT}, 
we go through a sequence of successive changes of integration variable,
first $y\to \widetilde{y}= y +\sqrt{1+y^2}$, 
next $\widetilde{y}\to \widehat{y} = \widetilde{y}^2-1$, 
finally $\widehat{y}\to u = \widehat{y}^{-2}-1$,
ending up with the elementary integral 
\newpage

\bea
  \int_\eps^{1/2\sqrt{2}}
\frac{f^\prime(y)}{y^2} 
   \dd{y}
&=&
-\int_0^{[2\eps(\eps+\sqrt{1+\eps^2})]^{-2}-1}
 \frac{\dd{u}}{\sqrt{u}}
\nonumber
\\
&=&
-\frac{1}{\eps}
\sqrt{\left(\eps+\sqrt{1+\eps^2}\right)^{-2} -4\eps^2}
=
-\frac{1}{\eps} + 1 + O(\eps)
\,.
\label{eq:EPSdownlimitINTevaluated}
\eea
 With \refeq{eq:EPSdownlimitINTevaluated}, the limit $\eps\downarrow{0}$ in
\refeq{eq:EPSdownlimitINT} gives us
\beq
  \int_0^{1/2\sqrt{2}}
\frac{1+f^\prime(y)}{y^2} 
  \dd{y}
=
1 -2\sqrt{2} 
\,,
\label{eq:COULtermASYMPint}
\eeq
which together with \refeq{eq:AnullOFsONEasympNEXTtoLEADfinalINT} proves 
\refeq{eq:AnullOFsONEasymptoticsNEXTtoLEAD}. 

 As for the term 
$
U(\beta/|{\SPvec{s}}_1|) 
= 
1- |{\SPvec{s}}_1|(A_0({\SPvec{s}}_1)-A_{\mathrm{Born}}^{(-)}({\SPvec{0}}) )
$
in \refeq{eq:AatSoneASYMPTOTICS}, we write it 
as the difference of the r.h.s.\refeq{eq:AnullOFsONEasympNEXTtoLEADfinalINT}
and the expression between big parentheses on the l.h.s. 
of \refeq{eq:AnullOFsONEasympNEXTtoLEADfinalINT}, 
\beq
U(\zeta)
=
  \int_0^{2\sqrt{2}}
\!\!
\left(
1-\frac{1}
     {\sqrt{1+ \zeta^4\xi^4}}
\right)
\dd\xi
+
  \int_0^{1/2\sqrt{2}}
\left(
1-\frac{1}
     {\sqrt{1+ \zeta^4/y^4}}
\right)
\frac{1+f^\prime(y)}
     {y^2} 
\dd{y},
\label{eq:UofBETAoverSone}
\eeq
then
use property (i) for $f^\prime$, then the convexity estimate 
$1- \sqrt{1+ \zeta^4\xi^4}^{\, -1} \leq \haelfte  \zeta^4\xi^4$
and drop a negative term to estimate 
\bea
|U(\zeta)|
&\leq& 
  \int_0^{2\sqrt{2}}
\!\!
\left(
1-
\frac{1}
     {\sqrt{1+ \zeta^4\xi^4}}
\right)
\dd\xi
+
  \int_0^{1/2\sqrt{2}}
\left(
1-
\frac{1}
     {\sqrt{1+ \zeta^4/y^4}}
\right)
\left|
\frac{1+f^\prime(y)}
     {y^2} 
\right|
\dd{y}
\nonumber
\\
&\leq&
\frac{\zeta^4}{2}
\left(
  \int_0^{2\sqrt{2}}
\!\!
\xi^4
\dd\xi
-
  \int_\zeta^{1/2\sqrt{2}}
\frac{1+f^\prime(y)}{y^6} 
\dd{y}
\right)
-
\int_0^{\zeta}
\frac{1+f^\prime(y)}{y^2} 
\dd{y}
\,.
\label{eq:ERRORtermASYMP}
\eea
 The remaining integrals in \refeq{eq:ERRORtermASYMP} containing $f^\prime$ evaluate to
\bea
-
\frac{\zeta^4}{2}
  \int_\zeta^{1/2\sqrt{2}}
\!\!
\frac{1+f^\prime(y)}{y^6} 
\!\!\!&\dd{y}&\!\!\!
-
\int_0^{\zeta}
\frac{1+f^\prime(y)}{y^2} 
\dd{y}
\nonumber
\\
&=&
\frac{3}{2}\zeta +2\zeta^2 + O(\zeta^3) 
+ 
\frac{1}{4}\zeta +\zeta^2 + O(\zeta^3) 
\eea
where we used property (iv) and the fact that 
by formula \refeq{eq:fDEFINITION} for $f$ 
singularities occur only at $y=\pm{i}$ and $y=1/2\sqrt{2}$.
 Finally, 
\beq
  \int_0^{2\sqrt{2}}
\!\!
\xi^4
\dd\xi
=
\frac{1}{5}\left(2\sqrt{2}\right)^5 
\,.
\eeq
 Adding up the numbers and rounding up, and recalling that $\zeta=\beta/|\SPvec{s}_1|$
and that $1+f^\prime <0$, we find
\beq
- 2\textstyle\frac{\beta}{|\SPvec{s}_1|}+O\left(\textstyle\frac{\beta^2}{|\SPvec{s}_1|^2}\right) 
< 
U\left(\textstyle{\frac{\beta}{\SPvec{s}_1}}\right)
<
20
\textstyle{\frac{\beta^4}
       {|\SPvec{s}_1|^4}}
\,.
\eeq
 
 Finally, we sharpen the bound from below on $U$ by noting that
for  $\zeta = \beta/|\SPvec{s}_1|\ll 1$, we have 
$1- \sqrt{1+ \zeta^4\xi^4}^{\, -1} = \haelfte  \zeta^4\xi^4 + O(\zeta^8)$, 
while the integral containing $1+f^\prime$ in \refeq{eq:UofBETAoverSone} 
is $O(\zeta)$; hence 
\beq
U(\zeta)
=
- \zeta\frac{3}{2}
  \int_0^\infty
\left(1 - \frac{t^2}{\sqrt{1+t^4}}\right)
\dd{t}
+ O(\zeta^2)
\,.
\label{eq:UASYMP}
\eeq
 Hence, $U(\beta/|\SPvec{s}_1|) < 0$ for $|\SPvec{s}_1|$ big enough. 

 This concludes the proof of our Proposition \ref{AnullASYMPTOTICS}.
\hfill \QED

 We believe that Proposition \ref{AnullASYMPTOTICS} yields the first rigorous 
and explicit results about the electrostatic two-body problem for 
the nonlinear Maxwell--Born--Infeld field equations.
 More important is the fact that the asymptotic expansion 
\refeq{eq:AatSoneASYMPTOTICS}
of $A_0(\SPvec{s}_1)$ for $|\SPvec{s}_1|\to\infty$ given in Proposition 
\ref{AnullASYMPTOTICS} verifies that the theory produces 
the correct Coulomb law for the dependence on $|\SPvec{s}_1|$ of the electrostatic
configurational energy of two point charges which are far apart.
 We remark that if we would have obtained a different 
asymptotic power law than $1/|\SPvec{s}_1|$ for 
$
A_0({\SPvec{s}}_1)-A_{\mathrm{Born}}^{(-)}({\SPvec{0}}) 
$, 
or the correct power law but with a coefficient 
different from unity, the theory would
not be able to reproduce the known physical data correctly.
 While the asymptotic expansion \refeq{eq:AatSoneASYMPTOTICS} of $A_0(\SPvec{s}_1)$ 
is therefore a gratifying result to have, 
upon reflection it is actually somewhat surprising that it is true at all! 
--- for it means that as $|\SPvec{s}_1|\to\infty$, 
the nonlinearity of the Maxwell--Born--Infeld field equations shows in 
the leading, but not in the next-to-leading order term of the asymptotic 
expansion, only to show again in the next-to-next-to-leading order. 
 This seems like a curious behavior for a nonlinear field theory.

 The next step is to show that for the two-body solution $\,{^\sharp}A_0(\SPvec{s},\SPvec{s}_1)$
of \refeq{eq:FOLIeqAstaticHYDRO}, the stationary Klein--Gordon equation
\beq
-\Delta_1 \psi^{\mathrm{stat}}(\SPvec{s}_1) 
+
\left(1 - |\veps + \alpha A_0(\SPvec{s}_1)|^2\right)\psi^{\mathrm{stat}}(\SPvec{s}_1) 
=
0
\,
\label{eq:KGeqHYDROGEN}
\eeq
with asymptotic condition $\psi^{\mathrm{stat}}({\SPvec{s}_1})\to 0$ for $|{\SPvec{s}_1}|\to\infty$
admits bound states.
 As the most immediate spin-off of Proposition \ref{AnullASYMPTOTICS} we indeed have

\begin{Coro}
\label{boundSTATES} 
 The Klein--Gordon equation \refeq{eq:KGeqHYDROGEN}
has infinitely many bound states.
\end{Coro}

\noindent
\textit{Proof of Corollary \ref{boundSTATES}}. 
 Rewriting the stationary Klein--Gordon equation \refeq{eq:KGeqHYDROGEN}
as
\beq
- \haelfte\Delta_1 \psi^{\mathrm{stat}}(\SPvec{s}_1) 
- \left(\overline{\veps}\alpha \overline{A}_0(\SPvec{s}_1)
 + \haelfte \alpha^2 \overline{A}_0(\SPvec{s}_1)^2\right)
\psi^{\mathrm{stat}}(\SPvec{s}_1) 
=
\haelfte\left(-1 + \overline{\veps}^2\right)\psi^{\mathrm{stat}}(\SPvec{s}_1) 
\,,
\label{eq:KGeqHYDROGENasSCHROEDGReq}
\eeq
with $\overline{A}_0(\SPvec{s}_1) = {A}_0(\SPvec{s}_1)-A_{\mathrm{Born}}^{(-)}({\SPvec{0}})$
and $\overline{\veps} = \veps + {\alpha}A_{\mathrm{Born}}^{(-)}({\SPvec{0}})$, we see that on
the left-hand side of \refeq{eq:KGeqHYDROGENasSCHROEDGReq} we have a  Schr\"odinger operator with
Schr\"odinger potential 
$V_{\overline\veps}(\SPvec{s}_1) 
= - \overline{\veps}\alpha \overline{A}_0(\SPvec{s}_1)
- \haelfte \alpha^2 \overline{A}_0(\SPvec{s}_1)^2$. 
 Hence we can interpret \refeq{eq:KGeqHYDROGENasSCHROEDGReq} as a
Schr\"odinger eigenvalue problem with a one-parameter family of potentials
$V_{\overline\veps}$ together with the constraint that the eigenvalues $E$ have to be of the form 
$E = \haelfte\left(-1 + \overline{\veps}^2\right)$.
 Now, by Proposition \ref{AnullASYMPTOTICS}, $V_{\overline\veps}$
is the sum of a Rollnik potential and an arbitrarily small bounded potential.
 Furthermore, again by Proposition \ref{AnullASYMPTOTICS}, 
we have $V_{\overline\veps}(\SPvec{s}_1) \leq - C |\SPvec{s}_1|^{-2+\delta}$
for $|\SPvec{s}_1| > R_0$ and all $\overline\veps>0$.
 Hence, by Theorem XIII.6a of
         \cite{reedsimonBOOKiv}
we conclude that for any $\overline\veps>0$ the  
Schr\"odinger operator on the l.h.s. of \refeq{eq:KGeqHYDROGENasSCHROEDGReq}
has infinitely many negative eigenvalues $E_k(\overline\veps)$, $k\in\NN$.
 Furthermore, since $-\haelfte\Delta +  V_{\overline\veps}$ 
in \refeq{eq:KGeqHYDROGENasSCHROEDGReq} is a compact perturbation of $-\haelfte\Delta$, 
its infinite discrete spectrum accumulates at zero.
 Moreover, by a simple monotonicity argument we have 
$E_k(\overline\veps)\downarrow$ as $\overline\veps\uparrow$, 
while $\haelfte\left(-1 + \overline{\veps}^2\right)\uparrow$ as 
$\overline\veps\uparrow$; in particular 
$\haelfte\left(-1 + \overline{\veps}^2\right)$ ranges from $-\haelfte$ to $0$
as $\overline\veps$ moves from $0^+$ to $1$. 
\hfill \QED

 Two remarks: 
 First, one should be able to prove that the lowest eigenvalue of 
\refeq{eq:KGeqHYDROGEN} in fact corresponds to the 
radiation-free ground state of hydrogen.
 For this purpose one has to discuss the manifestly positive energy functional 
 which we defer to a separate publication.
 Second, the  normalization of the 
stationary $\psi^{\mathrm{stat}}$ follows from eq.\refeq{eq:QrhoOFpsi}; 
however, for the spectral calculations we may set
$\|\psi^{\mathrm{stat}}\|_{L^2} = 1$.

	\subsubsection{The discrete spectrum (estimating $\beta$)}

 Proposition \ref{AnullASYMPTOTICS} says that the value of the total 
electrostatic potential at the position of the point electron varies
with the distance $|\SPvec{s}_1|$ from the nucleus according to Coulomb's 
law when the electron is `far away from the nucleus.'
 The error bound on the asymptotic expansion \refeq{eq:AatSoneASYMPTOTICS} 
tells us more specifically how far away from the nucleus the electron 
needs to be so that Coulomb's law holds within any given relative error. 
 This error bound together with the impressive range of validity of Coulomb's law 
will translate into some upper bound for $\beta$ through comparison with the 
well-known spectral data for hydrogen. 
 The assumption of an infinitely massive nucleus does not invalidate these arguments.

\newpage

 Now, we do have a fully special-relativistic theory, but we 
have not incorporated spin. 
 It would thus be foolish to aim at a comparison 
with the relativistic details of the quantum mechanical hydrogen 
spectrum as computed from Dirac's equation with purely Coulombic 
potential 
		 \cite{glimmjaffeBOOK, Keppeler}
(not to speak of the fine details caused by the Lamb shift,
                \cite{glimmjaffeBOOK}).
 Yet, it seems reasonable to demand that the $\beta$-induced corrections
to the hydrogen spectrum as given to leading order in $\alpha$, 
i.e. $O(\alpha^2)$, by Schr\"odinger's equation with purely Coulombic potential, 
should be sub-dominant to the known $O(\alpha^4)$ 
corrections computed from Dirac's equation with purely Coulombic potential.
 
 To expand \refeq{eq:KGeqHYDROGENasSCHROEDGReq} to leading order in 
$\alpha$ we make the Ansatz $\psi^{\mathrm{stat}} = \psi_n + \delta\psi^{\mathrm{stat}}$,
with $|\delta\psi^{\mathrm{stat}}| \ll 1$, and $\overline\veps = 1 + E_n$, with $|E_n| \ll 1$,
and discard all terms that are subordinate to the respective leading order terms
(this also means to discard the $\alpha^2\overline{A}_0^2(\SPvec{s}_1)$ 
term in the Schr\"odinger potential). 
 This leads to 
\beq
- \haelfte\Delta_1 \psi_n(\SPvec{s}_1) 
- \alpha \overline{A}_0(\SPvec{s}_1)
\psi_n(\SPvec{s}_1) 
=
E_n \psi_n(\SPvec{s}_1) 
\,.
\label{eq:SCHROEDINGEReqHYDROGEN}
\eeq
 Here, $n$ refers to the $n$th bound state, of which there are infinitely many
by repeating the arguments of Corollary \ref{boundSTATES}; we do not display the degeneracy 
of the spectrum.
 Next, to estimate $\beta$, notice that only $\overline{A}_0(\SPvec{s}_1)$ enters in 
\refeq{eq:SCHROEDINGEReqHYDROGEN}, and since by Proposition \ref{AnullASYMPTOTICS} 
and its ensuing remark we have 
$\overline{A}_0(\SPvec{s}_1) =  |\SPvec{s}_1|^{-1}(1 - U(\beta/|\SPvec{s}_1)|)$, 
with $U(\beta/|\SPvec{s}_1)|)$ bounded and satisfying 
$|U(\beta/|\SPvec{s}_1|)| \leq 2 \beta/|\SPvec{s}_1|$ for small $\beta/|\SPvec{s}_1|$,
 for \emph{all} $|\SPvec{s}_1|$
we may now  take the limit $\beta\downarrow{0}$ in which 
$\overline{A}_0(\SPvec{s}_1) \to  |\SPvec{s}_1|^{-1}$, obtaining 
the Schr\"odinger equation for the Coulomb Hamiltonian 
\beq
- \haelfte\Delta_1 \psi_n^{(0)} (\SPvec{s}_1) 
- \alpha |\SPvec{s}_1|^{-1}\psi_{n}^{(0)} (\SPvec{s}_1) 
=
E_n^{(0)}  \psi_{n}^{(0)} (\SPvec{s}_1) 
\,
\label{eq:SCHROEDINGEReqCOULOMB}
\eeq
with 
\beq
 E_n^{(0)}  = - \textstyle{\frac{\alpha^2}{2(n+1)^2}},\qquad n = 0,1,2,...
\,
\eeq
(The indexing is chosen to meet the convention that $E_0$ means the ground state energy.)
 We pause for a moment to remark that our calculation so far rigorously vindicates the 
identification of $\alpha$ with Sommerfeld's fine structure constant.

 Next, applying standard first-order perturbation theory 
              \cite{reedsimonBOOKiv},
we can write
$\psi_n(\SPvec{s}_1) = \psi_{n}^{(0)} (\SPvec{s}_1) + \delta\psi_{n}^{(\beta)}(\SPvec{s}_1)$
and $E_n = E_n^{(0)} + \delta E_n^{(\beta)}$ to compute the $\beta$-induced correction 
$\delta E_n^{(\beta)}$ to $E_n^{(0)}$ to first order in $U$ as
\beq
\delta E_n^{(\beta)}
= 
 \alpha \int_{\RR^3} |\psi_n^{(0)}(\SPvec{s}_1)|^2
|\SPvec{s}_1|^{-1} U(\beta/|\SPvec{s}_1|) \dvol(\SPvec{s}_1) 
\,.
\label{eq:firstPERTURBenergy}
\eeq
 We only estimate the ground state energy correction, i.e. \refeq{eq:firstPERTURBenergy} 
with $n=0$, for which
\beq
 \psi_0^{(0)}(\SPvec{s}_1) 
= 
\sqrt{{\alpha^3}/{\pi}}
\; e^{-\alpha|\SPvec{s}_1|}
\,
\eeq
is the familiar normalized eigenfunctions of the ground state for \refeq{eq:SCHROEDINGEReqCOULOMB}.
 Proposition \ref{AnullASYMPTOTICS} and the remark thereafter (also note that  \refeq{eq:UofBETAoverSone}
yields $|U|\leq 2\sqrt{2}$) now give
\bea
&&\abs{\delta E_0^{(\beta)}} 
\leq
\alpha 2\sqrt{2}\!\!\int_{\{|\SPvec{s}_1|< R\}}^{}\!\!
|\psi_0^{(0)} (\SPvec{s}_1)|^2|\SPvec{s}_1|^{-1} \dvol(\SPvec{s}_1)
\nonumber
\\
&&\qquad\qquad\qquad\qquad
+
2K\alpha\beta\!\! \int_{\{|\SPvec{s}_1|\geq R\}}^{} \!\! |\psi_0^{(0)} (\SPvec{s}_1)|^2
|\SPvec{s}_1|^{-2}\dvol(\SPvec{s}_1)
\,,
\eea
for $K\geq 1$ big enough, and 
$R$ is to be chosen so that this expression is minimized.$^\dagger$
 For this purpose, we will assume that $\beta$ is not 
bigger than 1, vindicate this assumption a posteriori, then
bootstrap to a smaller estimate for $\beta$.
 Thus, assuming $\beta \leq 1$, we find $R = \beta {K}/\sqrt{2}$ (to leading order), 
which leads to the tentative estimate 
\beq
|{\delta E_0^{(\beta)}}|
\,\aprl\,
4K \alpha^3\beta
\,,
\label{eq:energyCORRECTIONinBETA}
\eeq
which, when
 $\beta\leq C \alpha$, is indeed of $O(\alpha^2)$ 
relative to the leading Coulomb term, which is $\propto \alpha^2$.
(Actually, \refeq{eq:energyCORRECTIONinBETA} should be an upper bound
for ${\delta E_0^{(\beta)}}$.)
 To bootstrap the estimate for $\beta$ further, 
note that the $\beta$-induced correction 
\refeq{eq:energyCORRECTIONinBETA} should
definitely not compete with the first relativistic Dirac--Coulomb 
correction to the non-relativistic Schr\"odinger-Coulomb spectrum, 
which enters as $\delta E_0^{DC} = \viertel\alpha^4$.
 Hence, with  $K= 1$ for concreteness, 
we need to have at least $\beta \leq 10^{-1}\alpha$; 
equality would result in roughly
equal contributions by both corrections. 
 However, since the Dirac correction gives excellent agreement, we can 
actually be confident that such a conservative estimate for the 
$\beta$-induced correction is too feeble and can be improved significantly.
 If a $\beta$-induced correction of not more than about 
one-hundredth of the first Dirac correction is allowed,
\refeq{eq:energyCORRECTIONinBETA} would give $\beta \leq 10^{-3}\alpha$, roughly.
 Less aggressively, since \refeq{eq:energyCORRECTIONinBETA} is merely an 
upper estimate, $\beta \leq C\alpha$ with $C$ less than 10 is still 
conceivable.$^\dagger$
 Born's value for $\beta$ is compatible with these estimates, but
so is the temptingly speculative thought that, perhaps, $\beta = \alpha$. 
 In any event, it is also conceivable that better estimates with 
a complete model with spin will give $\beta \ll \alpha$, which would rule 
out Born's value of $\beta$, and at the same time put an end to the speculation 
whether $\beta=\alpha$.

 We finally translate these estimates into actual distances beyond which 
Coulomb's law is valid.
 By Proposition \ref{AnullASYMPTOTICS}, we have that 
$A_0(\SPvec{s}_1)-A_{\mathrm{Born}}^{(-)}({\SPvec{0}}) 
= |\SPvec{s}_1|^{-1}$ within $1\%$ 
\hfill
\smallskip
\footnoterule
{\footnotesize $^\dagger$ This sentence was slightly improved after the galley corrections.}
\newpage

\noindent
relative error when
$|\SPvec{s}_1|\geq 200\beta$ in leading order asymptotics;
cf. with the validity of the Taylor expansion of $A_0(\SPvec{s}_1)$ for 
$|\SPvec{s}_1|\leq 2\sqrt{2}\beta \approx 2.8 \beta$. 
 Recall that the Compton wave length of the electron is our unit of length;
relative to this unit, the so-called classical electron 
radius equals Sommerfeld's fine structure constant $\alpha$, while the 
Bohr radius equals $1/\alpha$. 
 A value of $\beta\approx\alpha$ would mean that Coulomb's law
for the pair potential  between point nucleus and point electron
is valid to within $1\%$ relative error down to a distance of about 
200 classical electron radii.

	\subsubsection{The hydrogen atom in small-velocities approximation}

 Different from the stationary state treatment, where the velocities
of all charges were identically zero, to obtain the formulas for the 
regime of non-zero but small velocities we here resort to non-rigorous 
but very plausible arguments; again, while non-rigorous, we see no 
reason why one  should not be able to make these arguments rigorous.

 Thus, we not only assume that radiation-reaction can be neglected; indeed
to get the leading order effects we may even \emph{assume} that magnetic 
effects can be neglected.
 That is, we assume that the electron moves so slowly that the electromagnetic
potential for the total electromagnetic fields equals, at the position of the electron,
the electrostatic Maxwell--Born--Infeld potential for the instantaneous configuration.
 For the remaining steps we adapt Dirac's prescriptions for the Dirac equation as
reproduced in conventional textbooks.\footnote{We alert the reader to the fact that some 
                                             prescriptions for the Klein--Gordon equation (with given fields) 
					     that one can find in some otherwise excellent 
					     textbooks are not correct.}
 Thus, to obtain the non-relativistic limit of the  Klein--Gordon equation
\refeq{eq:KleinGordonSINGLEelectronEQ} on single-electron configuration
space with an infinitely massive nucleus at the origin, we insert the Ansatz
\beq
 \psi(t,\SPvec{s}_1) = e^{-i \scriptstyle{\frac{5}{2}}t}
 \check\psi(t,\SPvec{s}_1) 
\label{eq:SINGLEelectronPSIfastSLOW}
\eeq
into \refeq{eq:KleinGordonSINGLEelectronEQ}.
 In \refeq{eq:SINGLEelectronPSIfastSLOW}, $\check\psi(t,\SPvec{s}_1)$ is slowly varying
in time as compared to $e^{-i \scriptstyle{\frac{5}{2}}t}$, having an even slower 
varying time derivative; the unusual looking $5/2$ is due to a term 
$\alpha{A}_{\mathrm{Born}}^{(-)}({\SPvec{0}}) = - 3/2$ (assuming  $\beta = \beta_{\mathrm{Born}}$).
 In \refeq{eq:KleinGordonSINGLEelectronEQ}, after factoring out $e^{-i \scriptstyle{\frac{5}{2}}t}$
the terms of the type $const. \check\psi$ then cancel out, and we are left with
\beq
\Big(-{\partial}^2
+ 2(1+  {\alpha}\overline{A}_0(\SPvec{s}_1)i\partial
+2\alpha\overline{A}_0(\SPvec{s}_1)+\alpha^2\overline{A}^2_0(\SPvec{s}_1)
+ \Delta_1\Big)\check\psi (t,\SPvec{s}_1) 
= 
0
\,.
\label{eq:KleinGordonSINGLEelectronEQnonrel}
\eeq
 Next, invoking the so-called singular perturbation theory we may neglect the second-order 
time derivative as small versus the first-order time derivatives; we also neglect the 
$O(\alpha)$ term versus the $O(1)$ term in the coefficient of the first-order time derivative, 
which is justified as long as the electron stays sufficiently far away from the nucleus 
--- with our estimates on $\beta$ this means farther than a few electron Compton wavelengths, 
which are distances a factor $10$ to $10^2$ smaller than the Bohr radius of the hydrogen atom; 
and we neglect the $O(\alpha^2)$ term versus the $O(\alpha)$ term in the Schr\"odinger potential. 
 Then $\check\psi \approx \Psi$, with $\Psi$ solving Schr\"odinger's equation for 
the hydrogen atom with Coulomb--Born--Infeld potential,
\beq
i\partial_t \Psi(t,\SPvec{s}_1) 
 = 
- \haelfte\Delta\Psi(t,\SPvec{s}_1) 
- \alpha \overline{A}_0(\SPvec{s}_1) \Psi(t,\SPvec{s}_1)
\,.
\label{eq:ERWINnucleusSINGLEelectronEQ}
\eeq
 Recalling our large distance asymptotics 
$\overline{A}_0(\SPvec{s}_1)\sim |{\SPvec{s}_1}|^{-1}$, which 
becomes exact for all $|\SPvec{s}_1|$ in the limit $\beta\downarrow 0$ and which 
holds very accurately in the range of validity of \refeq{eq:ERWINnucleusSINGLEelectronEQ},
we see that at the same level of accuracy we may now replace
\refeq{eq:ERWINnucleusSINGLEelectronEQ} by Schr\"odinger's
equation with the traditional Coulomb potential $-\alpha|{\SPvec{s}_1}|^{-1}$
as Schr\"odinger potential.

 Finally, in the same non-relativistic approximation, the relativistic guiding equation 
for the (negative) electron, \refeq{eq:dBBeq} with $\SPvec{v}^{\mathrm{qu}}$ given by 
\refeq{eq:QvelocityOFpsi}, reduces to the de Broglie--Bohm guiding equation
\beq
\bulldif{\SPvec{r}} (t)
=
\Im \left(\Psi^{-1} {\nabla}_1 \Psi\right)(t,\SPvec{r}(t))
\,.
\label{eq:deBroglieBohmGeq}
\eeq

             \section{On the de Broglie--Bohm guiding law}

 The Schr\"odinger equation \refeq{eq:ERWINnucleusSINGLEelectronEQ} (with the further
approximation $\overline{A}_0(\SPvec{s}_1)\approx |{\SPvec{s}_1}|^{-1}$) is of course 
well accepted as a basic equation of the non-relativistic quantum mechanics of the hydrogen 
atom (with infinitely massive nucleus). 
 The guiding equation \refeq{eq:deBroglieBohmGeq} is not. 
 It was first proposed by de Broglie
 \cite{deBroglieB}
(see also chpts. 6, 9, 10 in 
   \cite{deBroglieC})
who, however, did not pursue this lead any further until his idea was 
re-discovered and its merits explained in great detail by Bohm 
    \cite{BohmsHIDDENvarPAPERS},
upon which de Broglie himself returned to this approach
 \cite{deBroglieD}. 
 
 At the non-relativistic level, and generalized to the many-body situation, the corresponding
Schr\"odinger equation together with the corresponding many-body de Broglie--Bohm
guiding equation on configuration space, have been shown to provide an unorthodox 
yet entirely consistent and paradox-free formulation of non-relativistic spin-less 
quantum mechanics in terms of which all the usual measurement axioms of the conventional 
formulation can be {explained} as effective rules of procedure
	 \cite{BohmsHIDDENvarPAPERS, BellBOOK, duerretalA, Shelly} 
--- to the extent that can reasonably be expected from a non-relativistic theory.
 It has been extended to include spin via the many-body Pauli equation and a generalization 
of \refeq{eq:deBroglieBohmGeq} involving the Pauli spinors.

 Yet, Einstein for instance considered it  ``too cheap'' a trick to get rid
of the ``measurement problem,''\footnote{The collection of Bell's articles 
						    \cite{BellBOOK} 
				    is mandatory reading. A very good collection of publications by Bell
				    and almost everyone else about the measurement problem is 
					    \cite{wheelerzurekBOOK}.}
and while one can only speculate about Einstein's reasons for his assessment, 
many physicists have raised a similarly spirited objection, that the de Broglie--Bohm 
guiding equation for the point particle motions seemed to be `artificially appended' 
to the autonomous Schr\"odinger dynamics, apparently without feedback from the 
contemplated particle motions.
 To the extent that a feedback from the actual motion of the one and only real
configuration of the particles in the world is meant, such a feedback does not
exist, indeed. 
 Yet a hint of some feedback loop from generic motions of proper point particles
into the Schr\"odinger equation could logically have been seen in the expression 
for the Coulomb energy of proper point charge configurations that enters the 
Schr\"odinger equation for atoms and molecules. 
 In any event, as long as a consistent dynamical theory of point charges 
and electromagnetic fields was not available, this feedback loop remained speculative.
 Although explicitly shown here only for the hydrogen atom, the de Broglie--Bohm 
formulation of non-relativistic quantum mechanics with Coulomb Hamiltonian and autonomous 
Schr\"odinger dynamics obtains in the non-relativistic limit of our least invasively 
quantized relativistic electromagnetic field theory with point charges. 
 While in our formulation the actually real particles and fields configuration of the world
does not enter the system of Maxwell--Born--Infeld $^\sharp$field plus Klein--Gordon 
equations,\footnote{The initial conditions for the actual electromagnetic fields 
                    are inherited to some extent by the initial conditions for the 
		    $^\sharp$fields through the condition that these initial $^\sharp$fields 
		    become the actual initial fields when the actual initial particle configuration 
		    is substituted for the generic one.} 
in this system no dynamical equation is truly autonomous in itself, 
with $\psi$ providing the guiding field for the generic point charge sources of the 
$^\sharp$field equations, the solutions of which in turn providing the potentials with
in the Klein--Gordon equation. 
  In this sense a certain amount of feedback from the guiding field over the $^\sharp$
fields back into Schr\"odinger's equation exists. 
   Our present work thereby lends new support to the de-Broglie--Bohm formulation 
of non-relativistic quantum mechanics:
since in the relativistic theory the set of dynamical variables comes as a closed package 
from which no subset of it may be left out, its non-relativistic limit now shines a
new light on the de-Broglie--Bohm guiding equation in which it no longer appears as 
artificially appended to the Schr\"odinger equation.

	\section{Summary and Outlook}

	In this paper, we succeeded in the partial quantization of 
the UV problem-free classical theory of electromagnetism with
point charges that we developed in 
      \cite{KiePapI}.
  Here, `partial' refers to the fact that only the 
degrees of freedom of point charge motion are affected by the 
quantization procedure. 
 Our procedure is `least invasive' in the sense that it does not
tamper with the integrity of the classical mathematical structures 
which guarantee the absence of divergence problems for the classical theory. 
  We emphasize that at no point in our procedure have we replaced the mathematical 
objects of the classical theory by operators. 
 This guarantees that the mathematical integrity of the whole formalism is
left intact, which is what we mean by least invasive quantization.
 Singularities feature merely as mild defects in the electromagnetic potentials, 
and since no UV infinities are associated with them, no regularization of the 
defects and renormalization of parameters is called for. 

 The theory developed in 
 \cite{KiePapI}
and in this paper produces a relativistically covariant  actual electromagnetic 
spacetime, but the laws of motion require a foliation which has to be granted a 
certain reality of its own; cf. also 
  \cite{berndlETal, munchberndlETal, ShellyRodi}.
 The formalism should apply to the physics of positive or negative point electrons
to the extent that spin effects and the photonic nature of the electromagnetic fields 
can be neglected; however, the relativistic quantum theory is worked out explicitly 
here only for a single electron coupled to its fields. 

 In both the classical and the quantum theory, the actual electromagnetic fields 
are solutions of the Maxwell--Born--Infeld field equations with the actual point 
charges as sources, which move according to relativistic guiding laws.
 In the classical theory, the guiding law is of Hamilton--Jacobi type, 
generated by a solution of a relativistic Hamilton--Jacobi PDE coupled 
self-consistently to a configuration space indexed family of electromagnetic 
potentials satisfying corresponding field equations. 
 In the quantum theory, the guiding law for a single point charge reveals 
itself as of relativistic de Broglie--Bohm type, with the guiding 
field generated by a solution of a relativistic Klein--Gordon PDE
coupled to the configuration indexed field equations.
 The relativistic many electrons wave function formalism requires modifications 
that we briefly commented on; however, in the non-relativistic particles limit 
the many electrons formalism can be worked out along the lines of this paper.
 In that regime only minor modifications of the theory are needed to accommodate 
the electromagnetic effects of other, non-genuinely electromagnetic
particles, such as nuclei with or without magnetic moment and form factor,
spinning or not. 
	This requires putting in by hand the parameters 
$z_k$ for the charge number,  $\kappa_k$  for the ratio of the 
the electron's to the $k$-th nucleus' rest mass, and, if desired, 
a smeared-out spinning charge distribution.

 We remark that the dynamical equations of the quantum theory presented here 
are as deterministic as those of the classical theory, with a well-defined 
joint Cauchy problem for the wave function, the electromagnetic fields, and 
the point charges.
 In the non-relativistic limit these equations reduce to the dynamical equations 
of the de-Broglie--Bohm formulation of non-relativistic quantum mechanics,
for which its Cauchy problem with Coulomb and Newton interactions has
been proved to be globally well-posed
           \cite{berndlETAL}.
 It is therefore reasonable to expect that at least local well-posedness
of our relativistic Cauchy problem can be proved to hold; global well-posedness
is not necessarily to be expected in a relativistic theory. 

 Another rewarding aspect of our formalism is that the quantum theory does not require 
postulating any additional ``measurement axioms;'' more poetically
speaking, it does not suffer from ``the malaise of the measurement problem''
		\cite{wightmanPUB}.
 The dynamical equations of the theory themselves say what happens in a measurement,
in principle at least. 
 Whether the theory does make all the same predictions that the conventional relativistic
quantum formalism without spin and photon would make is to be doubted, at least we
do not see any reason why literally the same predictions should come out. 
 In any event, while our theory certainly achieves a certain consistent generalization of 
the de-Broglie--Bohm formulation of non-relativistic quantum mechanics with Coulomb 
interactions to the relativistic purely electromagnetic world, without the approximation 
of involving only given fields, it is a difficult open problem of exactly how much of the 
non-relativistic Bohmian quantum formalism can be consistently reconciled with relativity theory
  \cite{berndlETal, munchberndlETal, ShellyRodi}.

 As a by-product of finally having a dynamically consistent formulation 
of electrodynamics with point charge(s), we are now also able for the 
first time to address the problem of the correct value of Born's 
aether constant $\beta$, which enters the theory through the Born--Infeld 
laws of the aether. 
 We found in 
  \cite{KiePapI}
that Born's computation, based as it is on his dynamically incomplete
formulation of the theory, is inconclusive.
 Born calculated the value of $\beta$ by identifying the empirical electron 
rest mass ${\me}$ $({\times}c^{2})$ with the electrostatic energy of his 
spherically symmetric electrostatic solution for a single point charge.
 Our calculations here are based on the non-relativistic 
hydrogen spectrum, for which we worked out the first explicit and rigorous
results in the static two-body problem for the Maxwell--Born--Infeld
field equations.
 While our estimates for $\beta$ leave the value 
calculated by Born viable for now, the
definitive calculation of $\beta$ can be done only after spin,
and perhaps also the photon, have been incorporated into the theory.
 These are the issues that we will take up next.

\bigskip

\centerline{{EPILOGUE}}

{\small{

\ \ \ \ \ \
``I will say, though, that my feeling is that one of the 
  really unsolved problems is
\hfill

\ \ \ \  
to include the EM field. 
You have also banged your head against it. \ ..... \ \ This is 
\hfill

\ \ \ \  
not a criticism of the Bohm formalism. It is a statement that 
the problem is known 
\hfill 

\ \ \ \ 
(to some of us) to be serious and has to be addressed and it is not easy at all.''
 \hfill 

{\hskip 4truecm Elliott H. Lieb (private communication, Aug. 29, 2001)}
}}

 By implementing the notion of the point electron consistently into
the classical relativistic theory of electromagnetism and then, through 
least invasive quantization, into a spin- and photon-less quantum theory,
we also accomplished a consistent implementation of the total 
relativistic electromagnetic fields into the Bohm formalism of quantum theory.
 Yet these are only the first steps. 
 The incorporation of spin and the photon have to be addressed, 
and other burning issues, such as pair creation and annihilation, 
will hopefully be understood along these lines as well, in due course.
 Meanwhile I hope that Elliott H. Lieb, to whom this two-parts paper 
is dedicated in admiration on occasion of his 70th birthday, will take some 
pleasure in the fact that his comments and advice, offered at a very early 
stage of this work, have played a not unimportant r\^{o}le in its creation.
 Elliott has been a constant inspiration and encouragement, and I hope 
that he will continue to inspire and encourage us all for many many 
more years to come.
\bigskip

\textbf{Acknowledgments}:  This work began in early 1992 when the author held
a German-Dartmouth distinguished visiting professorship at Dartmouth College. 
 It was supported in the past two years by NSF grant DMS-0103808.
 I am indebted to many individuals, but I am most grateful to 
S. Goldstein and H. Spohn for many invaluable scientific 
discussions about electromagnetism and quantum theory.
 I also thank S. Chanillo for Moser's theorem, and J. Taylor for his
insights into the Fermi bundle.
 I owe very special thanks to S. Goldstein and, especially, to R. Tumulka 
for their helpful comments and penetrating criticisms of an earlier version 
of this paper, which prompted me to improve and clarify the presentation.
 My sincere thanks go also to the five referees for their favorable
reactions to this non-mainstream paper and their helpful suggestions.
 I thank T. Dorlas for sending me copies of Schr\"odinger's 
Dublin papers, and Y. Brenier and I. Bia{\l}ynicki-Birula for bringing
their more recent works to my attention after the first version of this 
paper was circulated.

\newpage


\scriptsize{

}
\end{document}